\documentclass[aps,prd,twocolumn,superscriptaddress,groupedaddress,preprintnumbers,showkeys,nofootinbib]{revtex4}  
\usepackage[latin1]{inputenc}                  
\usepackage{graphicx} 
\usepackage{dcolumn}  
\usepackage{bm}       
\usepackage{textcomp}
\usepackage{epsfig}
\usepackage{epsf}
\usepackage{bbm}

\usepackage{latexsym}                        
\usepackage{amsfonts}                           
\usepackage{amssymb}                            
\usepackage{amsmath}                            
\usepackage[mathscr]{eucal}                    
\usepackage{dcolumn}                             
\usepackage{multirow}                             
\usepackage{bm}                                 
\usepackage{hyperref}                         
\usepackage[usenames]{color}
\usepackage{array}
\usepackage{appendix}
\usepackage{bbold}
\usepackage{soul,xcolor}
\usepackage{lipsum}
\usepackage{braket}
\setstcolor{red}
 \usepackage{float}

\newcommand{\be}{\begin{equation}}
\newcommand{\ee}{\end{equation}}
\newcommand{\bea}{\begin{eqnarray}}
\newcommand{\eea}{\end{eqnarray}}

\def \3{\ss }

\newcommand{\tr}{\operatorname{Tr}}

\newcommand{\beq}{\begin{eqnarray}}
\newcommand{\eeq}{\end{eqnarray}}

\renewcommand{\arraystretch}{1.8}

\newcommand{\ord}{\mathcal{O}}
\newcommand{\opc}{\mathcal{O}^\mu}
\newcommand{\gfi}{\gamma_5}
\newcommand{\gam}{\gamma_\mu}

\newcommand{\jint}{\mathcal{J}}
\newcommand{\jintb}{\bar{\mathcal{J}}}

\newcommand{\gaumd}{g_A^{u-d}}
\newcommand{\gaupd}{g_A^{u+d}}
\newcommand{\gsumd}{g_S^{u-d}}

\usepackage{ulem}
\usepackage{letltxmacro}

\newif\ifcorrectingmode
\correctingmodetrue

\begin{document}

\title{Nucleon charges and $\sigma$-terms in lattice QCD}

\author{C.~Alexandrou} \affiliation{Department of Physics, University of Cyprus, P.O. Box 20537, 1678 Nicosia, Cyprus} \affiliation{Computation-based Science and Technology Research Center, The Cyprus Institute, 20 Kavafi Str., Nicosia 2121, Cyprus}
\author{S.~Bacchio} \affiliation{Computation-based Science and Technology Research Center, The Cyprus Institute, 20 Kavafi Str., Nicosia 2121, Cyprus}
\author{J.~Finkenrath} \affiliation{Department of Theoretical Physics, European Organization for Nuclear Research, CERN, CH-1211 Geneva 23, Switzerland}
\author{C.~Iona} \affiliation{Department of Physics, University of Cyprus, P.O. Box 20537, 1678 Nicosia, Cyprus}
\author{G.~Koutsou} \affiliation{Computation-based Science and Technology Research Center, The Cyprus Institute, 20 Kavafi Str., Nicosia 2121, Cyprus}
\author{Y.~Li} \affiliation{Department of Physics, University of Cyprus, P.O. Box 20537, 1678 Nicosia, Cyprus}
\author{G.~Spanoudes} \affiliation{Department of Physics, University of Cyprus, P.O. Box 20537, 1678 Nicosia, Cyprus}

\begin{abstract}

We  determine the nucleon axial, scalar and tensor charges and the nucleon $\sigma$-terms using twisted mass fermions. We employ three  ensembles with approximately equal physical volume of about 5.5~fm, three values of the lattice spacing, approximately  0.06~fm, 0.07~fm and  0.08~fm, and with the mass of the degenerate  up and down, strange and charm quarks tuned to approximately their physical values. We compute both isovector and isoscalar charges and $\sigma$-terms and their flavor decomposition including the disconnected contributions. We use the Akaike Information Criterion to evaluate  systematic errors due to excited states and the continuum extrapolation.  For the nucleon isovector axial charge   we find $g_A^{u-d}=1.250(24)$, in agreement with the experimental value.  Moreover, we extract the nucleon $\sigma$-terms and find for the light quark content $\sigma_{\pi N}=41.9(8.1)$~MeV and  for the strange $\sigma_{s}=30(17)$~MeV.
       \end{abstract}


\maketitle
\section{Introduction}
The nucleon axial, tensor and scalar charges, along with the $\sigma$-terms, are fundamental quantities within the Standard Model and provide insights into nucleon structure. The axial charge, $\gaumd$, determines the rate of neutron beta decay, providing a direct probe of chiral symmetry breaking in hadron physics. It also enters analyses of neutrinoless double-beta decay and plays a central role in tests of the unitarity of the Cabibbo-Kobayashi-Maskawa (CKM) matrix. Flavor-diagonal axial charges, $g_A^f$, describe the intrinsic spin, $ \frac{1}{2} \Delta \Sigma_q $, carried by quarks in the nucleon and are measured in deep inelastic scattering (DIS) experiments at facilities such as Jefferson Lab and CERN, with future plans at the Electron-Ion Collider (EIC).

The isovector tensor and scalar charges, $ g_T^{u-d} $ and $ g_S^{u-d} $, are less well known but are crucial in constraining beyond Standard Model (BSM) interactions~\cite{Bhattacharya:2011qm}. These charges provide essential theoretical input for interpreting results from ongoing neutrino scattering experiments, such as DUNE~\cite{Bischer:2018zcz}, COHERENT~\cite{COHERENT:2017ipa}, and GEMMA~\cite{Beda:2012zz}, as well as from direct dark matter detection searches~\cite{Undagoitia:2015gya,Chizhov:1995wp,Dhargyal:2016jgo}.  Providing an accurate determination of the tensor charge is important for phenomenological analyses of the transvesity parton  distribution function~\cite{Cocuzza:2023oam}.
The nucleon $\sigma$-terms measure  the contribution of quark masses to the nucleon mass, and enter in the determination of the elastic scattering cross section  between dark matter candidates, such as weakly interacting massive particles (WIMPs).

In this work, we compute the nucleon  charges and  $\sigma$-terms, using  twisted mass fermion ensembles at three different lattice spacings. These are simulated by the Extended Twisted Mass Collaboration (ETMC) using a mass degenerate up and down quark doublet, and the strange and charm quarks  ($N_f=2+1+1$) with all masses fixed to approximately their physical values, referred to as physical point. This allows for a precise determination of both isovector and flavor-diagonal charges, as well as the nucleon $\sigma$-terms, avoiding chiral extrapolations. Our results not only serve as a crucial benchmark for lattice QCD computations but also provide essential input for precision experiments aimed at probing BSM interactions, CP-violation, and dark matter detection.

\section{Methodology}
\label{sec:methodology}
The nucleon axial, tensor, and scalar  charges for each quark flavor $f$, denoted as $g^f_{\rm A,T,S}$, are derived from the nucleon matrix elements of the corresponding axial, tensor and scalar operators at zero momentum transfer. They are defined as

\begin{equation}
\langle N|\bar{\psi}^f\Gamma_{\rm A,S,T}\psi^f|N\rangle = g^f_{\rm A,S,T} \bar{u}_N\Gamma_{\rm A,S,T} u _N\,,
\end{equation}
where $u_N$ is the nucleon spinor and $\Gamma_A=\gamma_\mu\gamma_5$ for the axial-vector, $\Gamma_S=\mathbb{1}$ for the scalar and $\Gamma_T=\sigma_{\mu\nu}$ for the tensor operators. The renormalization group invariant $\sigma^f$-term is defined by $m_f\langle N|\bar{\psi}_f\psi_f|N\rangle$, with $m_f$ being the quark mass.

\subsection{Gauge ensembles and statistics}
\label{sec:stats}
To evaluate the nucleon matrix elements, we analyze gauge ensembles generated with  the twisted-mass fermion discretization scheme, which inherently offers $\ord(a)$-improvement~\cite{Frezzotti:2000nk,Frezzotti:2003ni}. A clover term is included in the action~\cite{Sheikholeslami:1985ij}, reducing isospin-breaking effects that arise from this fermion discretization. 

\begin{table}[h!]
	\centering
	{\small
		\renewcommand{\arraystretch}{1.2}
		\renewcommand{\tabcolsep}{1.5pt}
    \begin{tabular}{c|c|c|c|c|c}
    \hline \hline
        Ensemble     & Abrv. & $V/a^4$            & $\beta$ & $a$~[fm]    & $m_\pi$~[MeV] \\ \hline 
        cB211.072.64 & B64   & $64^3 \times 128$     & 1.778   & 0.07957(13) & 140.2(2)      \\ 
        cC211.060.80 & C80   & $80^3 \times 160$    & 1.836   & 0.06821(13) & 136.7(2)      \\ 
        cD211.054.96 & D96   & $96^3 \times 192$   & 1.900   & 0.05692(12) & 140.8(2)      \\ \hline \hline
        \end{tabular}}
    \caption{Parameters of the $N_f=2+1+1 $ ensembles analyzed in this work. In the first column, we give the name of
             the ensemble, in the second the abbreviated name, in the third the lattice volume, in the fourth $\beta = 6/g^2$ with $g$ the bare coupling constant, in the fifth
             the lattice spacing and in the last the pion mass. Lattice spacings and pion masses are
             taken from Ref.~\cite{ExtendedTwistedMass:2022jpw}.}
    \label{tbl:Ensembles}
\end{table}
The isosymmetric pion mass $m_\pi=$135 MeV~\cite{Alexandrou:2018egz,Finkenrath:2022eon} is reproduced through the tuning of the bare light quark parameter $\mu_l$. The parameters for the heavy quarks, $\mu_s$ and $\mu_c$, are tuned by utilizing the kaon mass along with a properly defined ratio between the D-meson mass and its decay constant, as well as a phenomenologically motivated ratio between the strange and charm quark masses, following the approach of Ref.~\cite{Alexandrou:2018egz,Finkenrath:2022eon}. The parameters of the ensembles used in this analysis are listed in Table \ref{tbl:Ensembles}. The lattice spacings and pion masses are adopted from Ref.~\cite{ExtendedTwistedMass:2022jpw}. Lattice spacing values are determined in both the meson and nucleon sectors, and we report the values from the meson sector, which agree with those from the nucleon mass as found in Ref.~\cite{ExtendedTwistedMass:2021gbo}.

\begin{widetext}

\begin{table}[h!]
\centering
\renewcommand{\arraystretch}{1}
\setlength{\tabcolsep}{3pt}
\begin{tabular}{|c|c|c|}
\hline
\multicolumn{3}{|c|}{\textbf{cB211.072.64}} \\
\multicolumn{3}{|c|}{749 configurations} \\
\hline
$t_s/a$ & $t_s$ [fm] & $n_{src}$ \\
\hline
8 & 0.64 & 1 \\
10 & 0.80 & 2 \\
12 & 0.96 & 5 \\
14 & 1.12 & 10 \\
16 & 1.28 & 32 \\
18 & 1.44 & 112 \\
20 & 1.60 & 128 \\
\hline
\multicolumn{3}{|c|}{Nucleon 2pt $n_{src}$=477} \\
\hline
\end{tabular}
\quad
\begin{tabular}{|c|c|c|}
\hline
\multicolumn{3}{|c|}{\textbf{cC211.060.80}} \\
\multicolumn{3}{|c|}{400 configurations} \\
\hline
$t_s/a$ & $t_s$ [fm] & $n_{src}$ \\
\hline
6 & 0.41 & 1 \\
8 & 0.55 & 2 \\
10 & 0.69 & 4 \\
12 & 0.82 & 10 \\
14 & 0.96 & 22 \\
16 & 1.10 & 48 \\
18 & 1.24 & 45 \\
20 & 1.37 & 116 \\
22 & 1.51 & 246 \\
\hline
\multicolumn{3}{|c|}{Nucleon 2pt $n_{src}$=650} \\
\hline
\end{tabular}
\quad
\begin{tabular}{|c|c|c|}
\hline
\multicolumn{3}{|c|}{\textbf{cD211.054.96}} \\
\multicolumn{3}{|c|}{494 configurations} \\
\hline
$t_s/a$ & $t_s$ [fm] & $n_{src}$ \\
\hline
8 & 0.46 & 1 \\
10 & 0.57 & 2 \\
12 & 0.68 & 4 \\
14 & 0.80 & 8 \\
16 & 0.91 & 16 \\
18 & 1.03 & 32 \\
20 & 1.14 & 64 \\
22 & 1.25 & 16 \\
24 & 1.37 & 32 \\
26 & 1.48 & 64 \\
\hline
\multicolumn{3}{|c|}{Nucleon 2pt $n_{src}$=480} \\
\hline
\end{tabular}
\end{table}

\begin{table}[h!]
\centering
\renewcommand{\arraystretch}{1}
\setlength{\tabcolsep}{2pt}
\begin{tabular}{|c|c|c|c|c|c|c|c|c|c|c|c|c|}
\hline
& \multicolumn{4}{c|}{\textbf{cB211.072.64}} & \multicolumn{4}{c|}{\textbf{cC211.060.80}} & \multicolumn{4}{c|}{\textbf{cD211.054.96}} \\

& \multicolumn{4}{c|}{749 configurations} & \multicolumn{4}{c|}{400 configurations} & \multicolumn{4}{c|}{494 configurations} \\
\hline
Flavour & $N_{defl}$ & $N_r$ & $N_{Had}$ & $N_{vect}$ & $N_{defl}$ & $N_r$ & $N_{Had}$ & $N_{vect}$ & $N_{defl}$ & $N_r$ & $N_{Had}$ & $N_{vect}$ \\
\hline
Light & 200 & 1 & 512 & 6144 & 450 & 1 & 512 & 6144 & 0 & 8 & 512 & 49152 \\
Strange & 0 & 2 & 512 & 12288 & 0 & 4 & 512 & 24576 & 0 & 4 & 512 & 24576 \\
Charm & 0 & 12 & 32 & 4608 & 0 & 1 & 512 & 6144 & 0 & 1 & 512 & 6144 \\
\hline
\end{tabular}
\caption{\small{Statistics used for the connected two- and three-point
    functions (top) and disconnected loops (bottom). Top: In each
    table, we provide the sink-source time separations in lattice
    units (first column) and physical units (second column) and the
    number of source positions per configuration (third column). For
    each ensemble, the bottom row indicates the number of source
    positions used for the two-point functions.} Bottom: For each
  ensemble, in the columns from left to right we give: i) the number
  of deflated eigenvectors $N_{defl}$, ii) the number of stochastic
  sources $N_r$, iii) the number of Hadamard vectors $N_{Had}$, and
  iv) the total number of computed vectors, $N_{vect}$, which after
  color and spin dilution are obtained via $12 \times N_ r \times
  N_{Had}$.}
\label{tbl:Stats}
\end{table}
\end{widetext}
To obtain the nucleon charges we evaluate two- and three-point nucleon
correlation functions and carry out fits to isolate the ground state
matrix element of interest as will be explained in detail in
Sec.~\ref{sec:FitProc}. In Table~\ref{tbl:Stats}, we present the
statistics used for the calculation of the correlation functions,
providing both the number of configurations analyzed and the number of
source positions per configuration. The three-point functions are
computed via the so-called \textit{fixed sink} method, where we carry
out a new inversion for each new sink-source separation. The number of
source positions per configuration listed in Table~\ref{tbl:Stats} are
increased with increasing separation so that the three-point function
statistical errors are maintained roughly constant.
Table~\ref{tbl:Stats} also includes details for the contributions from
disconnected quark loops. These are calculated for the light, strange,
and charm quark masses. To enhance the signal-to-noise ratio of these
disconnected loops, several noise-reduction techniques are
applied. These include the one-end trick~\cite{McNeile:2006bz}, exact
deflation of low modes~\cite{Gambhir:2016uwp}, and hierarchical
probing~\cite{Stathopoulos:2013aci}, as explained in Ref.~\cite{Abdel-Rehim:2016pjw} for the specific application to twisted mass fermions. The one-end trick is applied to
all loops for all three ensembles.  We also use hierarchical probing
for the loops computed for all quark loops with a probing distance of
8 for the light and strange quark loops and with 4 for the charm-quark
loops.  Deflation of low modes is applied to the light quark loops for
the cB211.072.64 and cC211.060.80 ensembles. For the cD211.054.96
ensemble a higher number of stochastic sources are used since using
deflation for this 96$^3\times$192 lattice, which scales with the
square of the volume, would be prohibitively expensive.

\subsection{Computation of correlators}
\label{sec:correlation functions}
Extracting the nucleon matrix elements involves the computation of
both three- and two-point Euclidean correlation functions. The
two-point function reads
\begin{equation}
    \begin{aligned}
        C(\Gamma_0,\vec{p};t_s) &= \sum_{\vec{x}_s}e^{-i(\vec{x}_s)\cdot \vec{p}} \times \\
                                    & \tr \left[ \Gamma_0 \braket{\jint_N(t_s,\vec{x}_s) \jintb_N (0,\vec{0}) } \right],
    \end{aligned}
\end{equation}
where we set the source at the origin, $t_s, x_s$ are the sink
coordinates, $\Gamma_0$ is the unpolarized, positive parity projector
$\Gamma_0=\frac{1}{2}(1+\gamma_0)$ and the interpolating field for the
nucleon
\begin{gather} \label{eq:IntField}
    \jint_N(t,\vec{x})=\epsilon_{abc}u_a(x) \left( u^T_b(x) C \gfi d_c(x) \right).
\end{gather}
In order to increase the overlap of the interpolating field with the
nucleon ground state and reduce contamination from excited states, so
that the ground state dominates for as small time separations as
possible, we apply \textit{Gaussian
  smearing}~\cite{Alexandrou:1992ti,Gusken:1989qx} to the quark fields
entering the interpolating field
\begin{gather}
    \widetilde{\psi}(\vec{x},t)=\sum_{\vec{y}}[1+a_G H(\vec{x},\vec{y};U(t))]^{N_G} \psi(\vec{y},t)
\end{gather}
where $\psi$ is either $u$ or $d$ of Eq.~(\ref{eq:IntField}) and $H$ is the hopping matrix 
\begin{gather}
    H(\vec{x},\vec{y};U(t))=\sum_{i=1}^{3}\left[ U_i(x)\delta_{x,y-\hat{i}}+U_i^\dagger(x-\hat{i})\delta_{x,y+\hat{i}} \right].
\end{gather}
The parameters $\alpha_G$ and $N_G$ are adjusted~\cite{Alexandrou:2018sjm,Alexandrou:2019ali} to achieve a nucleon smearing radius of approximately 0.5 fm. To suppress statistical noise arising from ultraviolet fluctuations, APE smearing~\cite{Albanese:1987ds} is applied to the gauge links used in the hopping matrix.

The three-point function for e.g.  the axial-vector  is given by
\begin{gather}
        C^\mu(\Gamma_k,\vec{q},\vec{p^\prime};t_s,t_{ins})=\sum_{\vec{x}_s,\vec{x}_{ins}} e^{i \vec{x}_{ins} \cdot \vec{q}} e^{-i \vec{x}_s \cdot \vec{p^\prime}} \times \notag \\ 
        \tr \left[ \Gamma_k \braket{\jint_N(t_s,\vec{x}_s) \opc(t_{ins},\vec{x}_{ins}) \jintb_N (0,\vec{0}) } \right],
\end{gather}
where $\Gamma_k=i \Gamma_0 \gfi \gamma_k$ is the polarized projector,
${\cal O}^\mu=\bar{\psi}\gamma_\mu\gamma_5\psi$ is the axial operator
insertion, $\vec{p^\prime}$ is the sink momentum and $\vec{q}$ is the
insertion momentum.

For the charges and $\sigma$-terms, we use $\vec{q}=0$ restricting to
the forward limit and to nucleons with no momentum boost,
i.e. $\vec{p^\prime}=0$. Similar expressions hold for the scalar and
the tensor three-point functions, with operator insertion ${\cal
  O}=\bar{\psi}\psi$ and ${\cal
  O}^{\mu\nu}=\bar{\psi}\sigma^{\mu\nu}\psi$, respectively. The
scalar matrix elements are extracted using the unpolarized projector
$\Gamma_0$, while the tensor requires the polarized, $\Gamma_k$. In
what follows we will use the notation for the axial-vector, keeping in
mind the straight forward generalization to the scalar and tensor
operator insertions.

\section{Analysis of correlators}
\label{sec:analysis}
The spectral decomposition of the two- and three-point functions 
are given respectively by 
\begin{gather}\label{eq:Twop}
    C(\Gamma_0,\vec{p};t_s)= \sum_{i}^{\infty} c_i (\vec{p})e^{-E_i(\vec{p})t_s}~~\text{and} \\
    C^\mu(\Gamma_k,\vec{q};t_s,t_{ins})=\sum_{i,j}^{\infty} A^{i,j}_\mu (\Gamma_k,\vec{q})e^{-E_i(\vec{0})(t_s-t_{ins})-E_j(\vec{q})t_{ins}} \label{eq:Threep}.
\end{gather}

The coefficients $c_i$ of the two-point function are overlap terms given by 
\begin{gather}
    c_i(\vec{p})=\tr[\Gamma_0 \braket{\jint_N|N_i(\vec{p})}\braket{N_i(\vec{p})|\jintb_N}]
\end{gather}
and the coefficients $A^{i,j}_\mu (\Gamma_k,\vec{q})$ in the three-point function are given by 
\begin{gather}
    A^{i,j}_\mu (\Gamma_k,\vec{q})=\tr[\Gamma_k \braket{\jint_N|N_i(\vec{0})} \braket{N_i(\vec{0})|\opc|N_j(\vec{q})} \notag  \\ 
    \braket{N_j(\vec{q})|\jintb_N}],
\end{gather}
where $\braket{N_i(\vec{0})|\opc|N_j(\vec{0})}$ is the matrix element
of interest for the axial case, and similarly the scalar and tensor
three-point functions give coefficients $A^{i,j} (\Gamma_0,\vec{q})$
and $A^{i,j}_{\mu\nu} (\Gamma_k,\vec{q})$, respectively. The overlaps
between the interpolating field and the nucleon state $\ket{N}$, such
as $\braket{\Omega|\jint_N|N}\equiv \braket{\jint_N|N}$, need to be
canceled in order for us to access the matrix element. In order to
cancel these unknown overlaps, we construct the ratio of the three- to
two-point function
\begin{gather}\label{eq:Ratio}
    R^\mu_A(\Gamma_k;t_s,t_{ins})=\frac{C^\mu(\Gamma_k,\vec{0};t_s,t_{ins})}{C(\Gamma_0,\vec{0};t_s)},
\end{gather}
i.e. setting $\vec{q}=\vec{0}$ in the three-point and $\vec{p}=\vec{0}$ in the
two-point functions.  In the limit of large time separations $\Delta E
(t_s-t_{ins})\gg 1$ and $\Delta Et_{ins}\gg1$, with $\Delta E$ being
the energy difference between the first excited state and ground
state, the ratio gives us the desired axial charge
\begin{gather}
    R^k_{A}(\Gamma_k;t_s,t_{ins}) \rightarrow g_{A} ,
\end{gather}
and similarly $R_{S}$ and $R_{T}^{\mu\nu}$ yield the scalar ($g_S$)
and tensor ($g_T$) charge respectively. Thus we use the parameters
$c_0$ and $A^{0,0}$ extracted from our data and substitute
Eqs.~(\ref{eq:Twop}) and (\ref{eq:Threep}) into Eq.~(\ref{eq:Ratio})
to obtain the charges via
\begin{align} \label{eq:ExtrRat}
 g_{S}=Z_P& \frac{{A^{0,0}}(\Gamma_0,\vec{0})}{c_0(\vec{0})},\nonumber\\
 g_{A}=Z_A&\frac{{A^{0,0}_k}(\Gamma_k,\vec{0})}{c_0(\vec{0})},\,\nonumber\\
g_{T}=Z_T&\epsilon^{ijk}\frac{{A^{0,0}_{ij}}(\Gamma_k,\vec{0})}{c_0(\vec{0})},
\end{align}
where $Z_P$, $Z_A$, and $Z_T$, are renormalization constants that will
be discussed in Sec.~\ref{sec:Renorm}. In practice, we need
to identify the smallest possible $t_s$ and $t_{ins}$ for which
excited state contributions are sufficiently suppressed. How fast
ground state dominance is achieved depends on the smearing procedure
applied on the interpolating fields and the energy gap between the
ground state and the excited states.  Additionally, noise increases
exponentially with $t_s$, so establishing convergence to the ground
state as early as possible is essential. We will employ a fitting
strategy of multi-state analysis of the contribution of the first
$N_{st}-1$ excited states. This aims to reliably determine the values
of $c_0$ and $A^{0,0}$ and is described in Sec.~\ref{sec:FitProc}.

\subsection{Fitting Strategy}\label{sec:FitProc}
To extract the ground state matrix elements, we perform simultaneous
fits to the two-point functions with the highest statistics and the
ratio of Eq.~(\ref{eq:Ratio}). To construct the ratio we divide the
three-point function with the two-point function having the same
statistics to take advantage of the correlations between them.

Since the optimal fit ranges in $t_s$ and $t_{ins}$ may vary for each case, we explore a wide parameter
space in the fitting ranges. In detail, we perform multiple fits by varying the following parameters:
\begin{itemize}
    \item $\bm{N_{st}}$: We truncate the sums over the energy states
      contributing to the two- (\ref{eq:Twop}) and three-point
      (\ref{eq:Threep}) functions to a maximum of $i_{max}=N_{st}-1$
      with $N_{st}=2$. We call these \textit{two-state} fits. If there
      is no detectable contamination due to excited states in the
      ratio, we fit it to a constant, performing a \textit{plateau}
      ($N_{st}=1$) fit.
    \item $\bm{t_{2pt}^{low}}$: We vary the lower time,
      $t_{2pt}^{low}$, in the fitting of the two-point functions,
      seeking for a region where only two states dominate.
    \item $\bm{t_s^{low}}$: Similarly, we vary the smallest value of
      the sink time $t_s$, we use for the fitting of the ratio and fit
      to all $t_s \geq t_s^{low}$ available.
    \item $\bm{t_{ins,0},t_{ins,s}}$: We vary the number of insertion
      time slices from the source and the sink times that we keep in
      the fits to the ratio. We use $t_{ins} \in
      [t_{ins,0},t_s-t_{ins,s}]$. For the charges we have $\vec{q}=0$
      so we fix $t_{ins,0}=t_{ins,s}$.
\end{itemize}

\subsection{Model Average}
From each combination of the varied parameters, we obtain a different result. We average 
the results using the \textit{Akaike Information Criterion (AIC)}~\cite{Jay:2020jkz,Neil:2022joj}.
In summary, for each fit $i$, we associate a weight $w_i$, which we define as 
\begin{gather}
    \log(w_i)=-\frac{\chi_i^2}{2}+N_{\text{dof},i},
\end{gather} 
where $N_{\text{dof}}=N_{\text{data}}-N_{\text{params}}$ is the number of degrees of freedom for each fit.
We then use the weights to define the probability
\begin{gather}\label{eq:Prob}
    p_i=\frac{w_i}{Z}~~\text{with}~~Z=\sum_{i}w_i.
\end{gather}
The \textit{Model Average (MA)} value of an observable $\ord$ is given as 
\begin{gather}
    \braket{\ord}_{MA}=\sum_{i} \bar{\ord}_i p_i~~\text{with}~~\sigma_{MA}^2=\sum_{i}(\sigma_{i}^2+\bar{\ord}^2_i)p_i - \braket{\ord}_{MA}^2
\end{gather}
with $ \bar{\ord}_i$ and $\sigma_{i}$ being the central value and error of the observable resulting 
from the $i_{th}$ fit.

\section{Renormalization}
\label{sec:Renorm}
Matrix elements computed in lattice QCD need to be renormalized in order to relate to physical observables. In the physical basis of the twisted-mass formulation, we use the renormalization functions $Z_A$ for the renormalization of the axial-vector current, $Z_P$ for the scalar and $Z_T$ for the tensor. We compute both flavor singlet and nonsinglet  renormalization factors for isoscalar and isovector combinations, respectively. We use gauge ensembles simulated specifically for the renormalization program with four mass degenerate quarks ($N_f=4$) at the same $\beta$ values as the three physical point ensembles used in the analysis of the matrix elements. For each $\beta$, we employ four ensembles of different sea quark masses to perform chiral extrapolations. We list these ensembles in Table~\ref{tbl:Zfact_ens}.

\begin{table}[h!]
\centering
{\normalsize
    \renewcommand{\arraystretch}{1.2}
    \renewcommand{\tabcolsep}{4.0pt}
\begin{tabular} {c | c | c | c | c | c}
  \hline \hline
   Ensemble & $\beta$  & $L^3 \times T$ & $a \mu_{\rm sea}$ & $\kappa$ & $c_{\rm SW}$ \\
   \hline
   cB4.060.24  & 1.778 & $24^3 \times 48$ & 0.0060 & 0.139305 & 1.6900 \\
   cB4.075.24  & 1.778 & $24^3 \times 48$ & 0.0075 & 0.139308 & 1.6900 \\
   cB4.088.24  & 1.778 & $24^3 \times 48$ & 0.0088 & 0.139310 & 1.6900 \\
   cB4.100.24  & 1.778 & $24^3 \times 48$ & 0.0100 & 0.139312 & 1.6900 \\
   \hline              
   cC4.050.32  & 1.836 & $32^3 \times 64$ & 0.0050 & 0.1386735 & 1.6452 \\
   cC4.065.32  & 1.836 & $32^3 \times 64$ & 0.0065 & 0.1386740 & 1.6452 \\
   cC4.080.32  & 1.836 & $32^3 \times 64$ & 0.0080 & 0.1386745 & 1.6452 \\
   cC4.095.32  & 1.836 & $32^3 \times 64$ & 0.0095 & 0.1386760 & 1.6452 \\
   \hline              
   cD4.040.48  & 1.900 & $48^3 \times 96$ & 0.0040 & 0.13793128 & 1.6112 \\
   cD4.050.48  & 1.900 & $48^3 \times 96$ & 0.0050 & 0.13793128 & 1.6112 \\
   cD4.065.48  & 1.900 & $48^3 \times 96$ & 0.0065 & 0.13793128 & 1.6112 \\
   cD4.080.48  & 1.900 & $48^3 \times 96$ & 0.0080 & 0.13793128 & 1.6112 \\
   \hline \hline
\end{tabular}
\caption{$N_f=4$ gauge ensembles  used for the computation of the renormalization functions. In the first column we give their name, in the second $\beta = (2 N/g^2)$, in the third the lattice volume ($L^3 \times T$) in units of $a$, in the fourth the twisted-mass parameter $(a \mu)$, in the fifth the  hopping parameter $\kappa$ and in the last the clover coefficient $c_{\rm SW}$.}
\label{tbl:Zfact_ens}
}
\end{table}

The renormalization factors are calculated nonperturbatively by employing the RI$'$/MOM scheme~\cite{Martinelli:1994ty} followed by perturbative conversion to $\overline{\rm MS}$ at the scale $\bar{\mu} = 2$ GeV. The RI$'$/MOM condition for the quark bilinear operators is 
\begin{equation}
{(Z_q^{{\rm RI}'})}^{-1} Z_{\Gamma}^{{\rm RI}'} \frac{1}{12} {\rm Tr} [\Lambda_{\Gamma} (p)(\Lambda_{\Gamma}^{\rm tree} (p))^{-1}] |_{p^2 = \mu_0^2} = 1,
\end{equation}
where $\Lambda_{\Gamma} (p)$ is the amputated vertex function of the bilinear operator with external quark fields, and $\Lambda_{\Gamma}^{\rm tree}$ is the corresponding tree-level value. $Z_q^{{\rm RI}'}$ is the renormalization factor of the quark field calculated using the following condition
\begin{equation}
Z_q^{{\rm RI}'} = \frac{1}{12} {\rm Tr} [S^{-1} (p) S^{\rm tree} (p)]= 1,     
\end{equation}
where $S(p)$ $(S^{\rm tree} (p))$ is the lattice (tree-level) quark propagator. The trace and the inverse in the above condition is taken over spin and color indices. The momentum $p$ of the vertex function and quark propagator is set to the renormalization scale $\mu_0$ of the RI$'$/MOM scheme.

We  employ the momentum source approach~\cite{Gockeler:1998ye}, in which the vertex functions are calculated with a momentum-dependent source. This method requires separate inversions for each value of the renormalization scale, but has the advantage of high statistical accuracy and the evaluation of the vertex for any operator at negligible computational cost. For the singlet operators, the vertex functions have an additional disconnected contribution, which is calculated using the same methods described in Sec.~\ref{sec:stats}, i.e. the one-end trick~\cite{McNeile:2006bz} and hierarchical probing~\cite{Stathopoulos:2013aci}.

The RI$'$ condition is defined for zero quark masses. To eliminate the mass contributions in the vertex functions, we  perform chiral extrapolations using the set of ensembles in Table~\ref{tbl:Zfact_ens} for each $\beta$ value. For $Z_A$ and $Z_T$, a linear fit with respect to the twisted-mass parameter $\mu_{\rm sea}$ is employed, given by
$$Z_\Gamma (\mu_0, \mu_{\rm sea}) = c_0 (\mu_0) + c_1 (\mu_0) \cdot (a \mu_{\rm sea}),$$ which is sufficient for removing the mild dependence on the quark mass, as observed in similar investigations, e.g., in Ref.~\cite{Alexandrou:2015sea}. The pseudoscalar vertex function suffers from a pion pole and a dedicated analysis is needed to remove it reliably. We follow the procedure described in Ref.~\cite{ExtendedTwistedMass:2021gbo}, where a non-unitary prescription is employed, in which the values of sea and valence quark masses are not equal. The pseudoscalar vertex functions are calculated for multiple values of the valence quark mass for each sea quark mass and a double chiral extrapolation is performed (see Ref.~\cite{ExtendedTwistedMass:2021gbo} for more details).

To reduce systematic errors related to lattice artifacts and rotational $O(4)$ breaking effects, we  employ spatially isotropic momenta of the form

$$(a p) \equiv 2 \pi [(n_t + 0.5)/(T/a), n_x/(L/a), n_x/(L/a), n_x/(L/a)],$$ where $n_i \in \mathbb{Z}$, $L \ (T)$ is the spatial (temporal) extend of the lattice, and we  employ the  momentum cuts given by  $\sum_{i} p_i^4/(\sum_i p_i^2)^2 < 0.3$. The momentum form respects the periodic (antiperiodic) boundary conditions applied on the quark fields at a spatial (temporal) direction. Also, an important aspect of our renormalization program is the improvement of the nonperturbative estimates by subtracting one-loop discretization errors, calculated in lattice perturbation theory. This procedure results in a milder dependence of the renormalization factors on $(a^2 p^2)$. Further details can be found in similar investigations of our group, see, e.g., Ref.~\cite{Alexandrou:2015sea}.

After chiral extrapolation and subtraction of one-loop artifacts, we  evolve the resulting values of the RI$'$/MOM renormalization factors at a large reference momentum scale $\mu_{\rm ref}$ using continuum perturbation theory:\footnote{The anomalous dimensions of the operators in RI$'$/MOM is known to 4 loops for nonsinglet operators ~\cite{Chetyrkin:1997dh,Gracey:2003yr,Gracey:2022vqr} and to 2 loops for singlet operators~\cite{Constantinou:2016ieh}.}
\begin{eqnarray}
    Z^{{\rm RI}'}_{\Gamma} (\mu_{\rm ref}^2,\mu_0^2) &=& Z^{{\rm RI}'}_{\Gamma} (\mu_0^2) \ \frac{R_{\Gamma}^{{\rm RI}'} (a_s(\mu_{\rm ref}^2))}{R_{\Gamma}^{{\rm RI}'} (a_s(\mu_0^2))}, \\
    R_{\Gamma}^{X} (x) &=& \exp \Big\{ \int^{x} dx' \frac{\gamma_{\Gamma}^{X} (x')}{\beta^{X} (x')}\Big\}
\end{eqnarray}
where $\gamma_\Gamma^X (a_s)$ is  the  anomalous dimension of the operator being considered and  $\beta^X (a_s)$ the $\beta$ function, within the scheme X. Note that the evolution of the scale is not applied in the nonsinglet axial-vector operator because it has zero anomalous dimension. In contrast, the singlet axial current has a nonzero anomalous dimension due to the axial anomaly~\cite{Larin:1993tq}. Then, we  apply a linear fit in $\mu_0^2$ to eliminate residual dependence on the initial scale $\mu_0^2$, of the form 
$$Z_\Gamma (\mu_{\rm ref}^2, \mu_0^2) = c_0 (\mu_{\rm ref}^2) + c_1 (\mu_{\rm ref}^2) \cdot \mu_0^2.$$

\begin{figure*}[ht!]
\includegraphics[width=\textwidth]{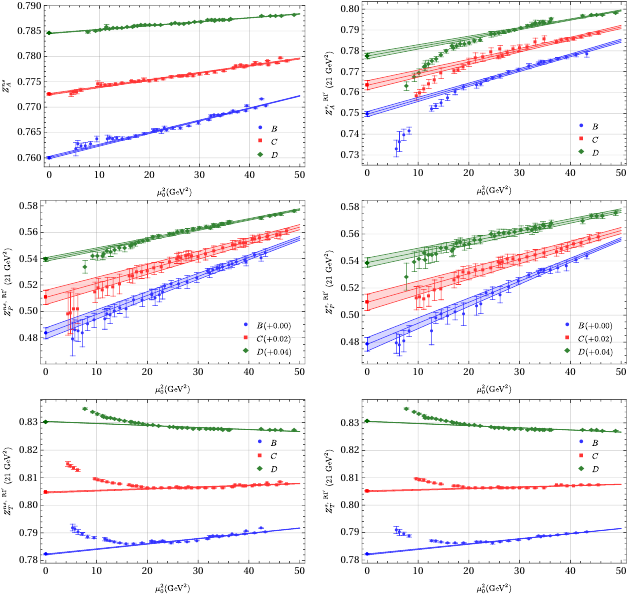} 
    \caption{Renormalization functions in RI$'$/MOM scheme at 21~GeV$^2$}
    \label{fig:Zfactors}
\end{figure*}

In Fig.~\ref{fig:Zfactors}, we provide the momentum fits for both singlet and nonsinglet renormalization factors $Z_P^{s (ns)}$, $Z_A^{s (ns)}$ and $Z_T^{s (ns)}$, for the three lattice spacings, using a reference scale of $\mu_{\rm ref}^2 = 21 {\rm GeV}^2$. The fit range is set to 12~GeV$^2$ - 50 GeV$^2$ for $Z_P^{s (ns)}$ and $Z_A^{ns}$, and 18~GeV$^2$ - 50 GeV$^2$ for $Z_A^{s}$ and $Z_T^{s (ns)}$. In these ranges, the linear fit can sufficiently describe the
residual momentum dependence. Lower momenta are excluded from the fits, where large nonperturbative effects (e.g., hadronic contaminations) are present and in which perturbative conversion does not work properly.

The extrapolated values of the renormalization factors at $\mu_0^2 = 0$ are converted to the $\overline{\rm MS}$ scheme at the reference scale of 2 GeV, using an intermediate Renormalization Group Invariant (RGI) scheme given by
\begin{equation}
Z_\Gamma^{\overline{\rm MS}} (4 {\rm GeV}^2) = C^{\overline{\rm MS}, {\rm RI}'}_\Gamma (4 {\rm GeV}^2, 21 {\rm GeV}^2) \cdot Z_\Gamma^{{\rm RI}'} (21 {\rm GeV}^2),
\end{equation}
where $C^{\overline{\rm MS}, {\rm RI}'}_\Gamma$ is calculated up to four loops (two loops) in perturbation theory for the nonsinglet (singlet) operators~\cite{Chetyrkin:1997dh,Gracey:2003yr,Gracey:2022vqr,Constantinou:2016ieh}. The final values for the renormalization factors of the quark bilinears under study in the $\overline{\rm MS}$ scheme and at 2 GeV are given in Table \ref{tbl:Renorm}. The error quoted in the parenthesis is the total error by adding in quadrature the statistical and systematic errors. The systematic error is estimated by varying the fit intervals in the momentum fits, as well as, by varying the highest perturbative order that we include in the analytical expression of evolution and conversion functions. 

\begin{table}[h!]
	\centering
	{\normalsize
		\renewcommand{\arraystretch}{1.2}
		\renewcommand{\tabcolsep}{4.0pt}
    \begin{tabular}{c|c|c|c}
    \hline \hline
                        & B64         & C80         & D96         \\ \hline
    $Z_A^{ns}$ & 0.7598(8) & 0.7724(4) & 0.7849(2) \\ \hline
    $Z_A^s$     & 0.7740(120)   & 0.7921(84)  & 0.8030(91)  \\ \hline
    $Z_P^{ns}$ & 0.4695(45) & 0.4771(55) & 0.4841(32) \\ \hline
    $Z_P^s$     & 0.4638(61)  & 0.4767(61)  & 0.4845(39)  \\ \hline
    $Z_T^{ns}$ & 0.8314(9) & 0.8516(33) & 0.8756(12) \\ \hline
    $Z_T^s$     & 0.8293(27)  & 0.8493(41)  & 0.8763(29)  \\ \hline \hline
    \end{tabular}
	}
	\caption{Renormalization factors for the singlet (s) and nonsinglet (ns) operators in the $\overline{\rm MS}$ scheme at 2 GeV.}
        \label{tbl:Renorm}
	\vspace*{-0.2cm}
\end{table}

\section{Results}
\label{sec:results}
\subsection{Nucleon charges}
\begin{figure}[ht!]
	\includegraphics[width=\columnwidth]{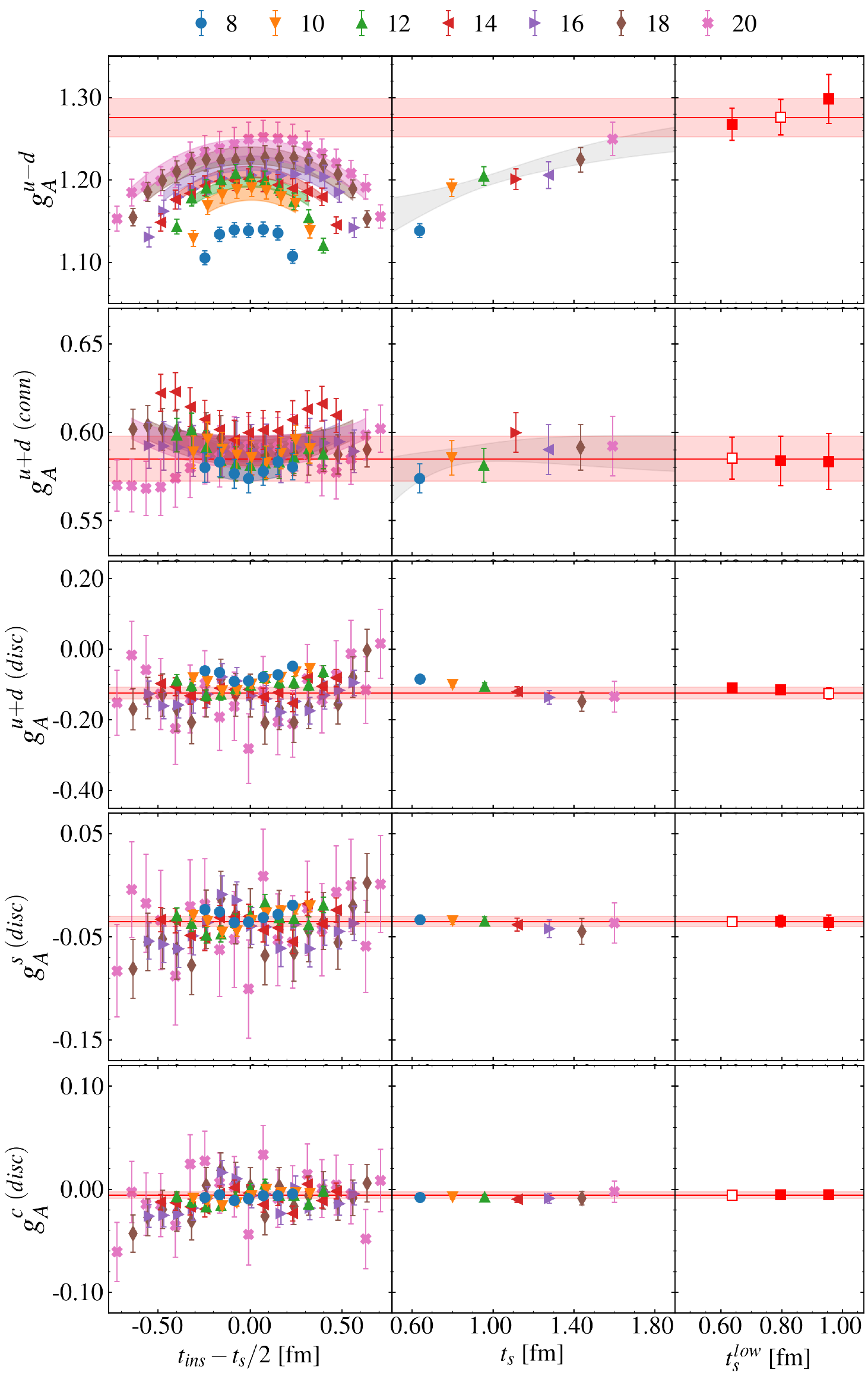}\\
	\vspace*{-0.3cm}
    \caption{Extraction of the isovector (first row), isoscalar
      connected (second row), isoscalar disconnected (third row),
      strange (fourth row), and charm (fifth row) contributions to the
      axial charge, for the B64 ensemble. In the first column, we show
      the ratios versus $t_{ins}-t_s/2$ with the symbol notation for
      each value of $t_s$ given in the legend. The horizontal bands
      that span all columns are the model averaged results.  In the
      middle column, we show the ratio versus $t_s$ for
      $t_{ins}=t_s/2$ when 2-state fits are performed (i.e. for the
      connected contributions) or the result of a constant fit on each
      $t_s$ separately when plateau fits are performed (i.e. for the
      disconnected contributions). When 2-state fits are performed, we
      also plot a gray band that shows the resulting ratio dependence
      on $t_s$ for $t_{ins}=t_s/2$ as predicted by the fit.  In the
      last column, we show the value of the fit with the largest
      weight for a fixed $t_s^{\rm low}$, as explained in the
      text. The open symbol shows the $t_s^{\rm low}$ with the largest
      weight. We note that for the disconnected data, odd source-sink
      separations are also used in the analysis, that are not plotted
      here for better visibility.}
	\vspace*{-0.4cm}
    \label{fig:gA_B64}
\end{figure}

\begin{figure}[ht!]
	\includegraphics[width=\columnwidth]{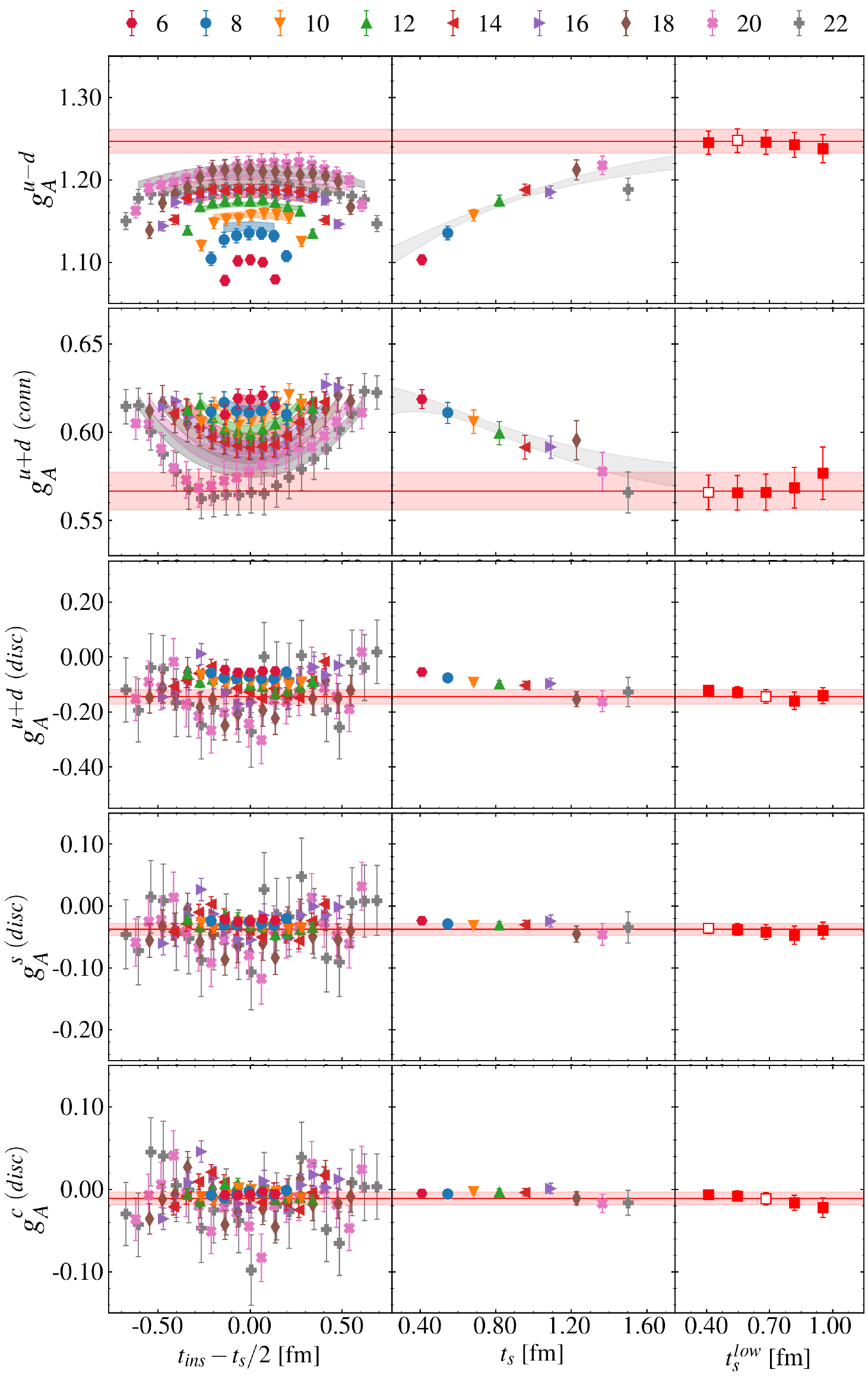}\\
	\vspace*{-0.3cm}
    \caption{Extraction of the axial charges for the C80 ensemble. The
      notation is the same as in Fig.~\ref{fig:gA_B64}. }
	\vspace*{-0.4cm}
    \label{fig:gA_C80}
\end{figure}

\begin{figure}[ht!]
	\includegraphics[width=\columnwidth]{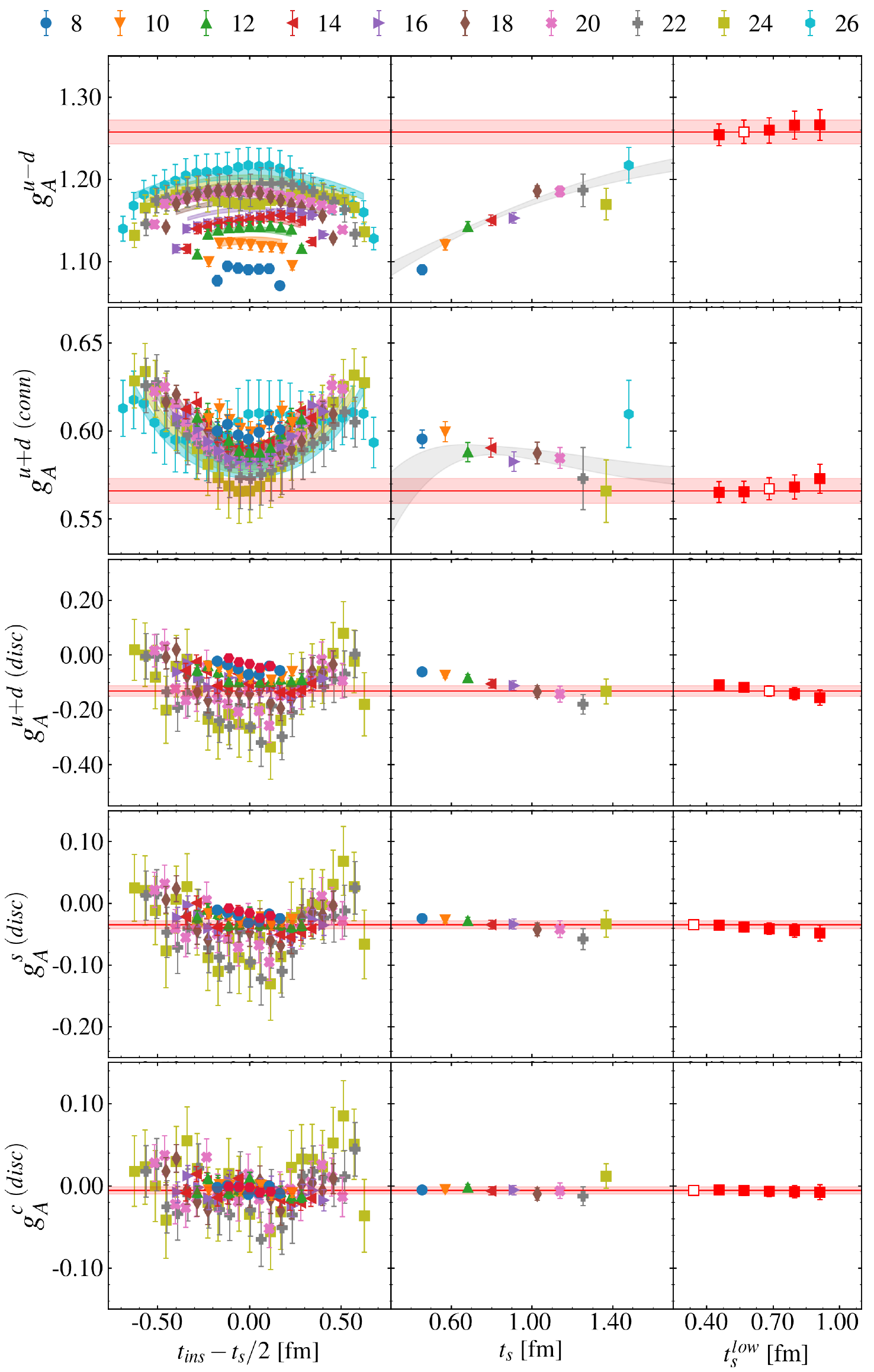}\\
	\vspace*{-0.3cm}
        \caption{Extraction of the axial charges for the D96 ensemble. The
      notation is the same as in Fig.~\ref{fig:gA_B64}. }
	\vspace*{-0.4cm}
    \label{fig:gA_D96}
\end{figure}

The analysis carried out for the extraction of the nucleon ground state matrix element for the axial charges is shown in Figs.~\ref{fig:gA_B64}, \ref{fig:gA_C80} and \ref{fig:gA_D96} for the three ensembles analyzed here. As mentioned previously, we obtain the desired charge from the matrix element of the appropriate current at zero momentum transfer. For example, the isovector axial charge is obtained by computing the three-point function of the axial-vector operator given by
\begin{gather}
    \ord^{u-d}_{A} \equiv  A_\mu = \bar{u} \gam \gfi u -\bar{d} \gam \gfi d,
\end{gather}
where $u$ and $d$ are the up and down quark fields respectively. 

As can be seen in Figs.~\ref{fig:gA_B64},~\ref{fig:gA_C80},
and~\ref{fig:gA_D96}, the ratios of the connected contributions, from
which the isovector and isoscalar combinations are determined, exhibit
a considerable dependence on $t_{ins}$ which indicates sizeable
excited state contributions. For these cases we therefore perform
2-state fits, following the fitting approach of
Sec.~\ref{sec:FitProc}, with $N_{st}=2$. Furthermore, for the
isovector case, we include in the fit the temporal component of the
three-point function with one unit of momentum transfer,
$C^0(\Gamma_k, \frac{2\pi}{L}\hat{k}; t_s, t_{ins})$, where $L$ is the
spatial length of the lattice and $\hat{k}$ a unit vector in the $k$
spatial direction, and correspondingly the two-point function with one
unit of momentum. This analysis is motivated by chiral perturbation
theory~\cite{Bar:2018xyi,Bar:2016uoj} which foresees an amplification
of the axial matrix element between a nucleon and pion-nucleon state,
i.e. the $A^{0,1}_\mu$ and $A^{1,0}_\mu$ coefficients. The temporal
component of the axial operator is obtained at no additional
computational cost in our setup and since it exhibits strong
dependence on the excited state can be used for a more precise
identification of the $\pi N$ energy and thus a better estimation of
the ground-state matrix element of interest.

For the disconnected contributions shown in
Figs.~\ref{fig:gA_B64},~\ref{fig:gA_C80}, and~\ref{fig:gA_D96}, namely
the isoscalar $\gaupd$ and the single-flavor $g_A^s$ and $g_A^c$, we
observe no detectable excited state contamination within the accuracy
of our results and we thus use plateau fits for these quantities. We
note that for the disconnected contributions, obtained by combining
the fermion loop with the two-point function, and thus in principle
have all values of $t_s$ available, we choose to plot even values of
$t_s$ for clarity and to match those of the connected contributions
(see Table~\ref{tbl:Stats}), where the values of $t_s$ need to be
chosen beforehand within the fixed sink approach employed here.

The maximum value of $t_s^{low}$ is taken to be approximately the same
in physical units (fm) for the three ensembles considered. This
corresponds to $t_s^{low}/a=12,\,14$ and $16$ or $t_s^{low}=0.96,
0.96\, {\rm and} \, 0.91$~fm for the B64, C80 and D96 ensembles,
respectively. For all three ensembles, we observe a very mild
dependence on $t_s^{low}$. The results in the right columns of
Figs.~\ref{fig:gA_B64}, \ref{fig:gA_C80} and \ref{fig:gA_D96},
correspond to the highest probability model for each value of
$t_s^{low}$, when varying $t_{{ins,0}}$ and $t_{2pt}^{low}$. Namely,
depending on the ensemble and quantity analyzed, for the connected
contributions where we perform 2-state fits, $t_{{ins,0}}$ varies from
$\sim$0.06 fm to $\sim$0.16 fm and $t_{2pt}^{low}$ varies from
$\sim$0.34~fm to $\sim$0.72~fm, while for the disconnected, where we
fit to a constant, $t_{{ins,0}}$ varies from $\sim$0.17 fm to
$\sim$0.40 fm.

We take the continuum limit, $a \rightarrow 0$, using the results from
the three available ensembles with different lattice spacings. We
carry out three types of extrapolation and evaluate a combined
systematic and statistical error via a model average over the three
fits. Namely, we use a linear fit in $a^2$ and a constant fit either
using all three ensembles or when omitting the coarser ensemble,
B64. A strong dependence on the lattice spacing will result in a model
average favoring the linear fit, while a mild $a^2$ dependence will
lead to a model average favoring the two constant fits.

\begin{figure}[ht!]
	\includegraphics[width=\linewidth]{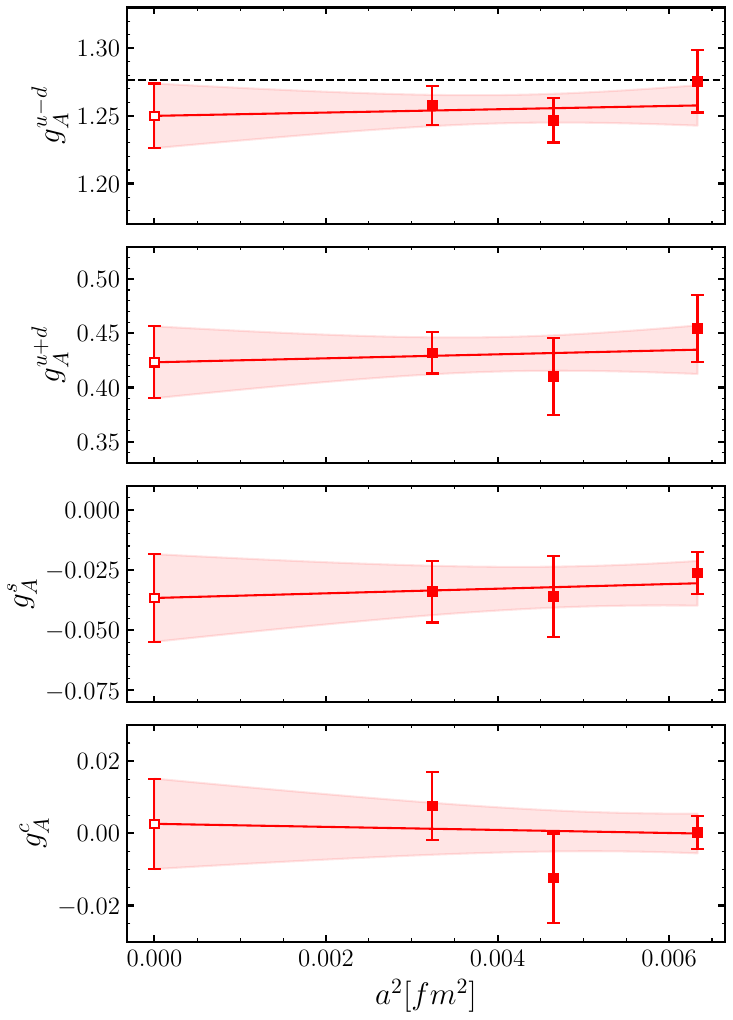}\\
	\vspace*{-0.3cm}
	\caption{Continuum limit of the nucleon axial charges (open
          symbols and band) extrapolated from three values of $a$
          (filled symbols). The extrapolation is the result of a model
          average which combines linear and constant fits as explained
          in the text. For the isovector $\gaumd$, the dashed line
          represents the experimental value~\cite{Markisch:2018ndu}.}
	\vspace*{-0.4cm}
    \label{fig:gA_extrap}
\end{figure}

The continuum limit extrapolations for the isovector, isoscalar, the
strange and charm axial charges are shown in
Fig.~\ref{fig:gA_extrap}. The isoscalar combination includes the
disconnected contributions. In Table~\ref{tbl:gA}, we give our results
for each gauge ensemble as well as our final continuum limit
extrapolated results.

\begin{table}[ht!]
	\centering
	{\footnotesize
		\renewcommand{\arraystretch}{1.2}
		\renewcommand{\tabcolsep}{2.0pt} 
		\begin{tabular}{c|c|c|c|c|c}
        \hline \hline
        $g_A$  & $u-d$     & $u+d$      & $u+d-2s$   & $u+d+s-3c$  & $u+d+s+c$  \\ \hline
        B64    & 1.275(23) & 0.454(31)  & 0.511(22)  & 0.432(20)   & 0.433(30)  \\ 
        C80    & 1.247(16) & 0.410(36)  & 0.489(28)  & 0.418(23)   & 0.368(65)  \\ 
        D96    & 1.258(14) & 0.432(19)  & 0.4952(95) & 0.371(21)   & 0.401(49)  \\ \hline
        $a= 0$ & 1.250(24) & 0.423(33)  & 0.490(20)  & 0.343(55)   & 0.382(70)  \\ \hline \hline
		\end{tabular}
        \\[0.1cm]
		\begin{tabular}{c|c|c|c|c}
        \hline \hline
        $g_A$  & $u$       & $d$        & $s$         & $c$        \\ \hline
        B64    & 0.867(18) & -0.408(14) & -0.0262(86) & 0.0003(45) \\ 
        C80    & 0.832(22) & -0.415(21) & -0.036(17)  & -0.012(12) \\ 
        D96    & 0.843(16) & -0.415(15) & -0.034(13)  & 0.0076(94) \\ \hline
        $a= 0$ & 0.832(28) & -0.417(22) & -0.037(18)  & 0.003(13)  \\ \hline \hline
		\end{tabular}
	}
	\caption{Values for the axial 2-, 3-, and 4-flavor isovector
          and isoscalar combinations (top) and the extracted single
          flavor charges (bottom) for each ensemble and in the
          continuum limit, using the model average strategy described
          in the text.}
	\label{tbl:gA}
	\vspace*{-0.2cm}
\end{table}

\begin{figure}[ht!]
	\includegraphics[width=\linewidth]{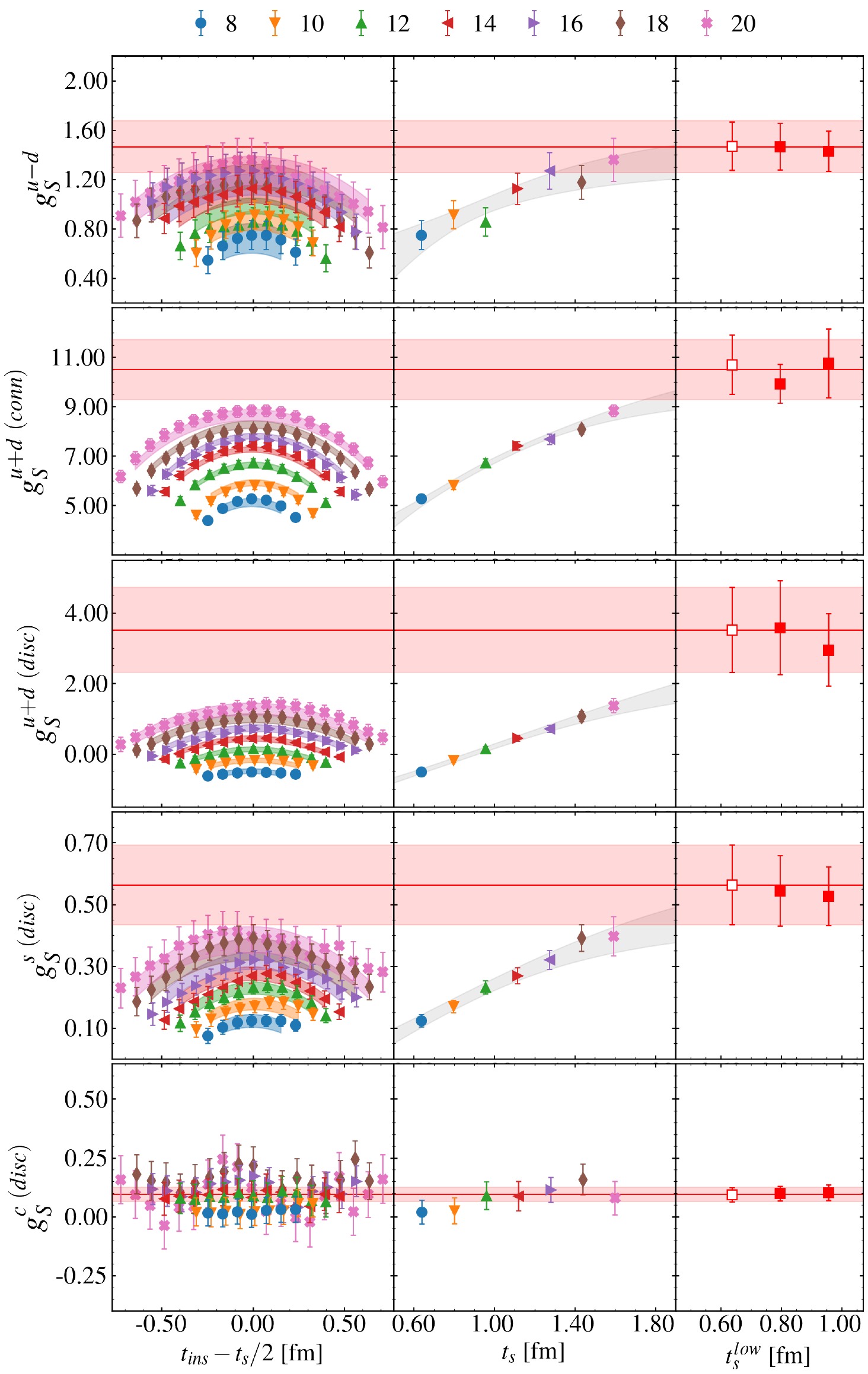}\\
	\vspace*{-0.3cm}
    \caption{The ratio and fit results for the B64 ensemble for scalar
      charges. The notation is the same as in Fig. \ref{fig:gA_B64},
      with 2-state fits being used for all cases except for $g_S^{c}$,
      where we use constant fits.}
	\vspace*{-0.4cm}
    \label{fig:gS_B64}
\end{figure}

\begin{figure}[ht!]
	\includegraphics[width=\linewidth]{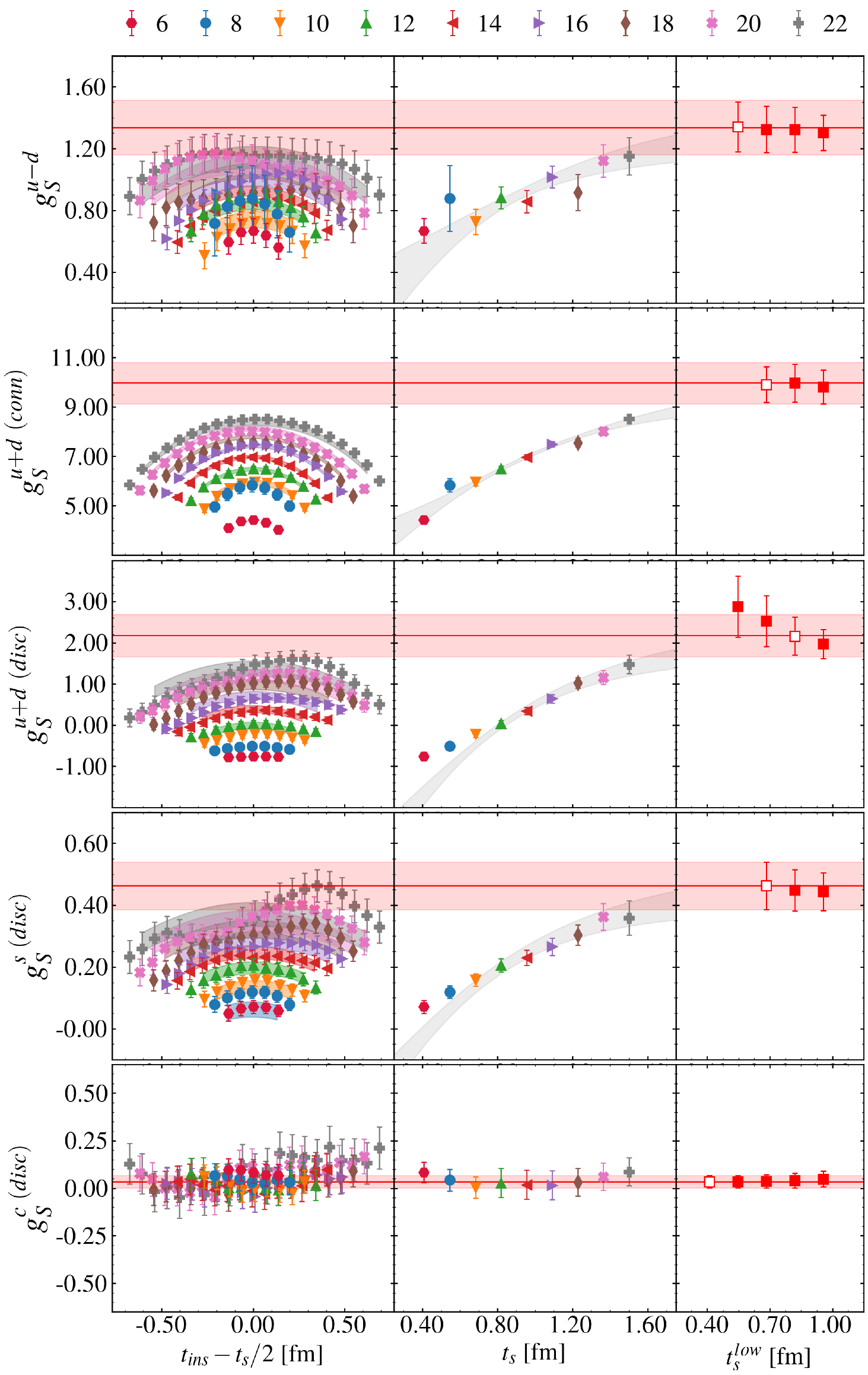}\\
	\vspace*{-0.3cm}
    \caption{The ratio and fit results for the C80 ensemble for scalar
      charges. The notation is the same as in Fig. \ref{fig:gA_B64}
      and the fits used for each charge are the same as in
      Fig. \ref{fig:gS_B64} }
	\vspace*{-0.4cm}
    \label{fig:gS_C80}
\end{figure}

\begin{figure}[ht!]
	\includegraphics[width=\linewidth]{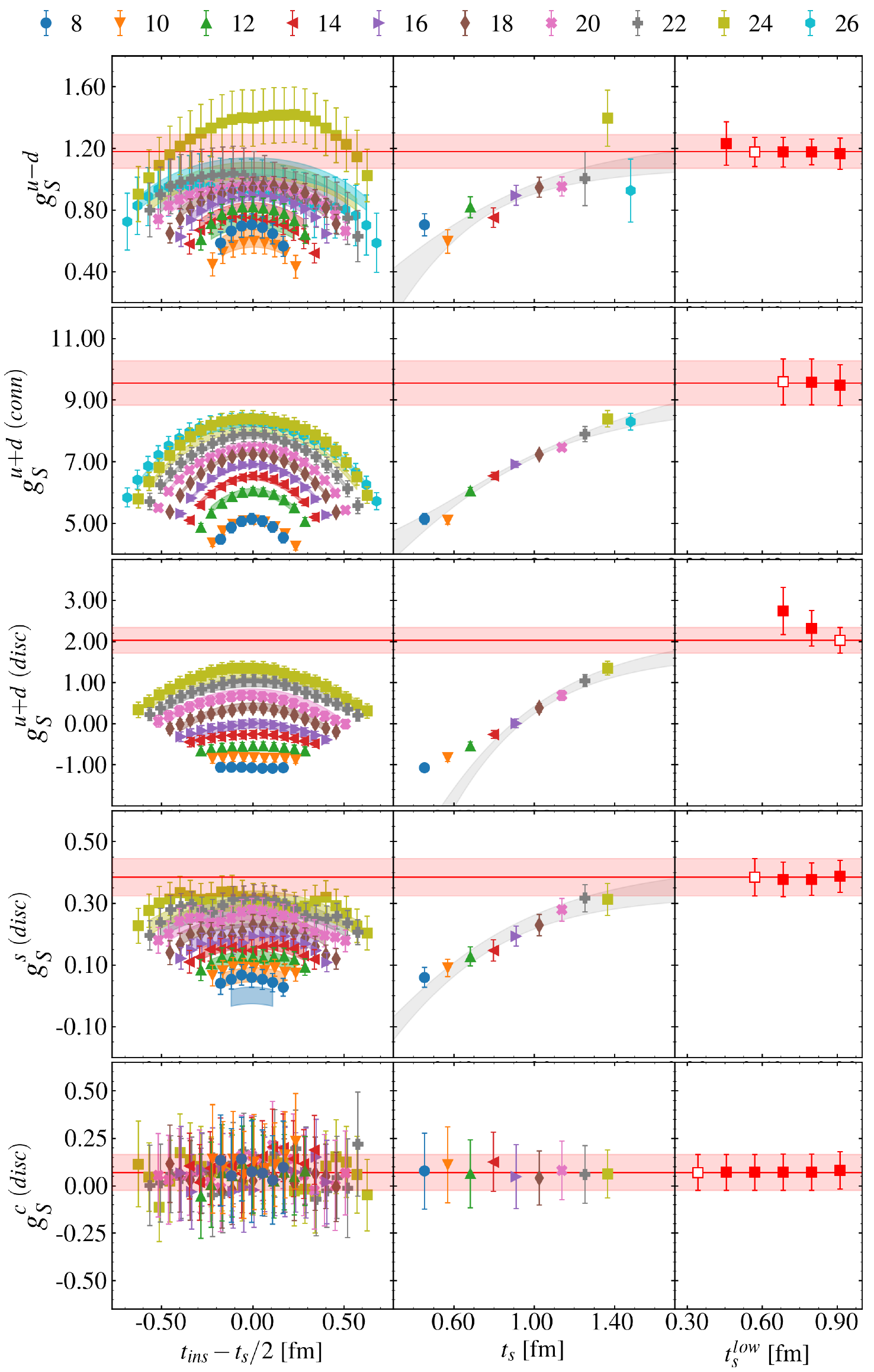}\\
	\vspace*{-0.3cm}
    \caption{The ratio and fit results for the D96 ensemble for scalar
      charges. The notation is the same as in Fig. \ref{fig:gA_B64}
      and the fits used for each charge are the same as in
      Fig. \ref{fig:gS_B64} }
	\vspace*{-0.4cm}
    \label{fig:gS_D96}
\end{figure}

\begin{figure}[ht!]
	\includegraphics[width=\linewidth]{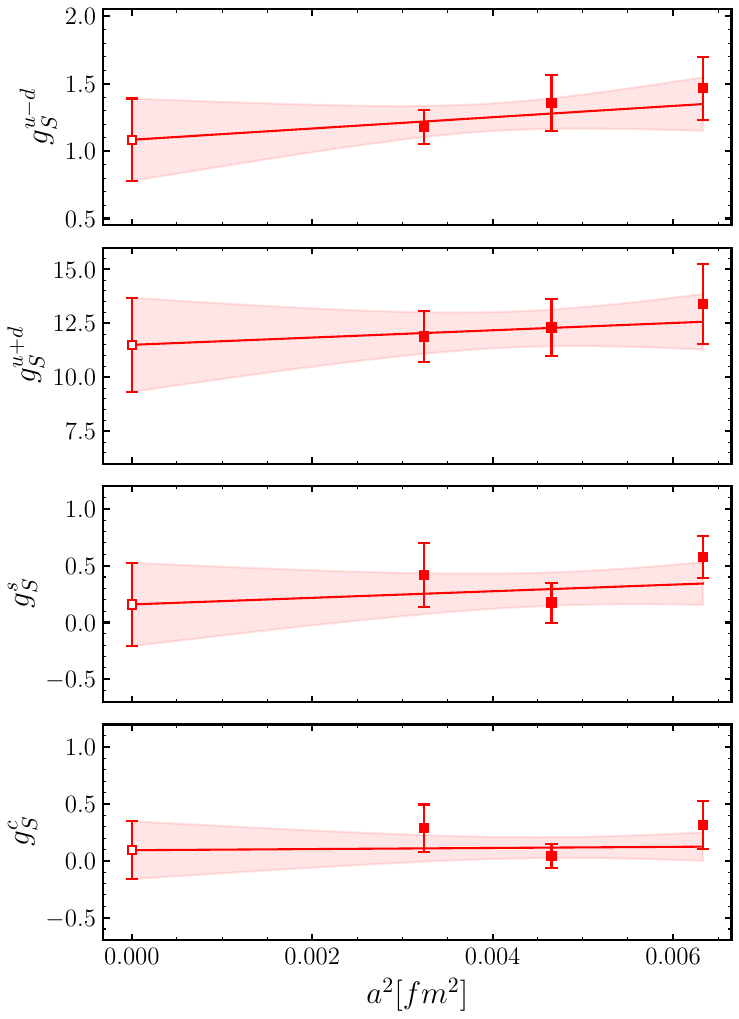}\\
	\vspace*{-0.3cm}
	\caption{Continuum limit of the nucleon scalar charges
          following the notation and procedure described in
          Fig.~\ref{fig:gA_extrap} for the axial case.}
	\vspace*{-0.4cm}
    \label{fig:gS_extrap}
\end{figure}

We perform the same analysis for the scalar charges as we did for the
axial. In absence of any insight similar to the axial case for the
dependence on excited states, we restrict to using ratios and
two-point functions at zero momentum. The ratios and extracted values
are shown in Figs.~\ref{fig:gS_B64}, \ref{fig:gS_C80} and
\ref{fig:gS_D96} for the three ensembles. Here we observe excited
state contamination and thus employ 2-state fits in all contributions
apart from the disconnected ratios yielding $g_S^c$. We also observe a
larger relative uncertainty for these quantities.  The continuum limit
is taken in the same way as for the axial charges and our final
results are presented in Fig. \ref{fig:gS_extrap} and in Table
\ref{tbl:gS}.

\begin{table}[ht!]
	\centering
	{\footnotesize
		\renewcommand{\arraystretch}{1.2}
		\renewcommand{\tabcolsep}{2.0pt} 
		\begin{tabular}{c|c|c|c|c|c}
        \hline \hline
        $g_S$  & $u-d$     & $u+d$      & $u+d-2s$   & $u+d+s-3c$  & $u+d+s+c$  \\ \hline
        B64    & 1.47(23)  & 13.4(1.9)  & 11.8(1.8)  & 12.6(1.9)   & 13.9(1.9)  \\ 
        C80    & 1.35(21)  & 12.3(1.3)  & 11.9(1.4)  & 12.2(1.3)   & 12.4(1.2)  \\ 
        D96    & 1.18(13)  & 11.9(1.2)  & 11.2(1.4)  & 11.5(1.2)   & 12.7(1.4)  \\ \hline
        $a= 0$ & 1.08(31)  & 11.5(2.2)  & 11.2(2.1)  & 11.4(2.1)   & 12.2(2.2)  \\ \hline \hline
		\end{tabular}
        \\[0.1cm]
		\begin{tabular}{c|c|c|c|c}
        \hline \hline
        $g_S$  & $u$       & $d$        & $s$        & $c$        \\ \hline
        B64    & 7.23(97)  & 5.76(85)   & 0.58(18)   & 0.32(21)   \\ 
        C80    & 6.78(68)  & 5.43(59)   & 0.17(18)   & 0.05(11)   \\ 
        D96    & 6.59(67)  & 5.41(62)   & 0.42(28)   & 0.29(21)   \\ \hline
        $a= 0$ & 6.4(1.1)  & 5.30(98)   & 0.16(37)   & 0.09(26)   \\ \hline \hline
		\end{tabular}
	}
	\caption{Values for the scalar 2-, 3-, and 4-flavor isovector
          and isoscalar combinations (top) and the extracted single
          flavor charges (bottom) for each ensemble and in the
          continuum limit.}
	\label{tbl:gS}
	\vspace*{-0.2cm}
\end{table}

\begin{figure}[h!]
	\includegraphics[width=\linewidth]{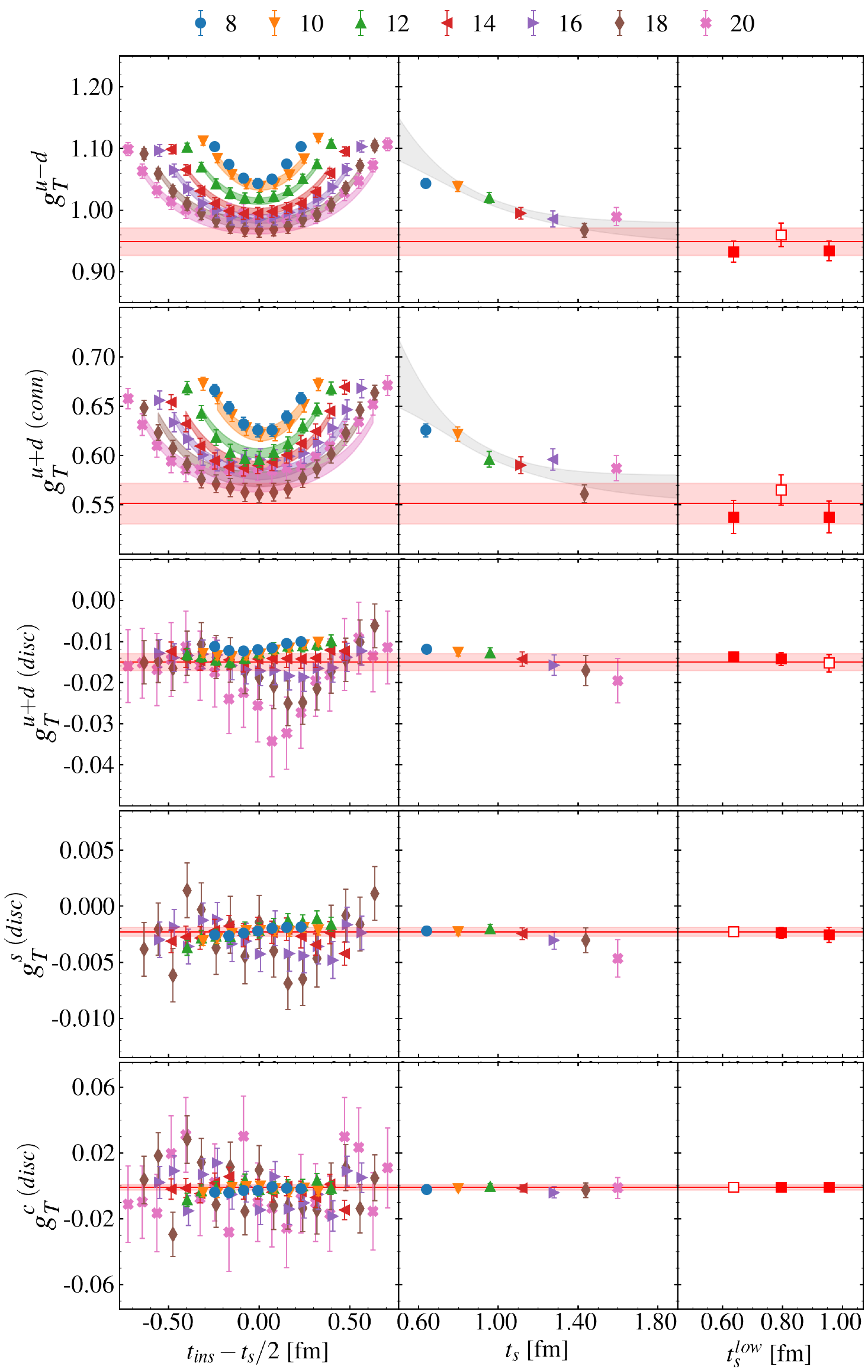}\\
	\vspace*{-0.3cm}
    \caption{The ratio and fit results for the B64 ensemble for tensor
      charges. The notation is the same as in Fig. \ref{fig:gA_B64}.}
	\vspace*{-0.4cm}
    \label{fig:gT_B64}
\end{figure}

\begin{figure}[h!]
	\includegraphics[width=\linewidth]{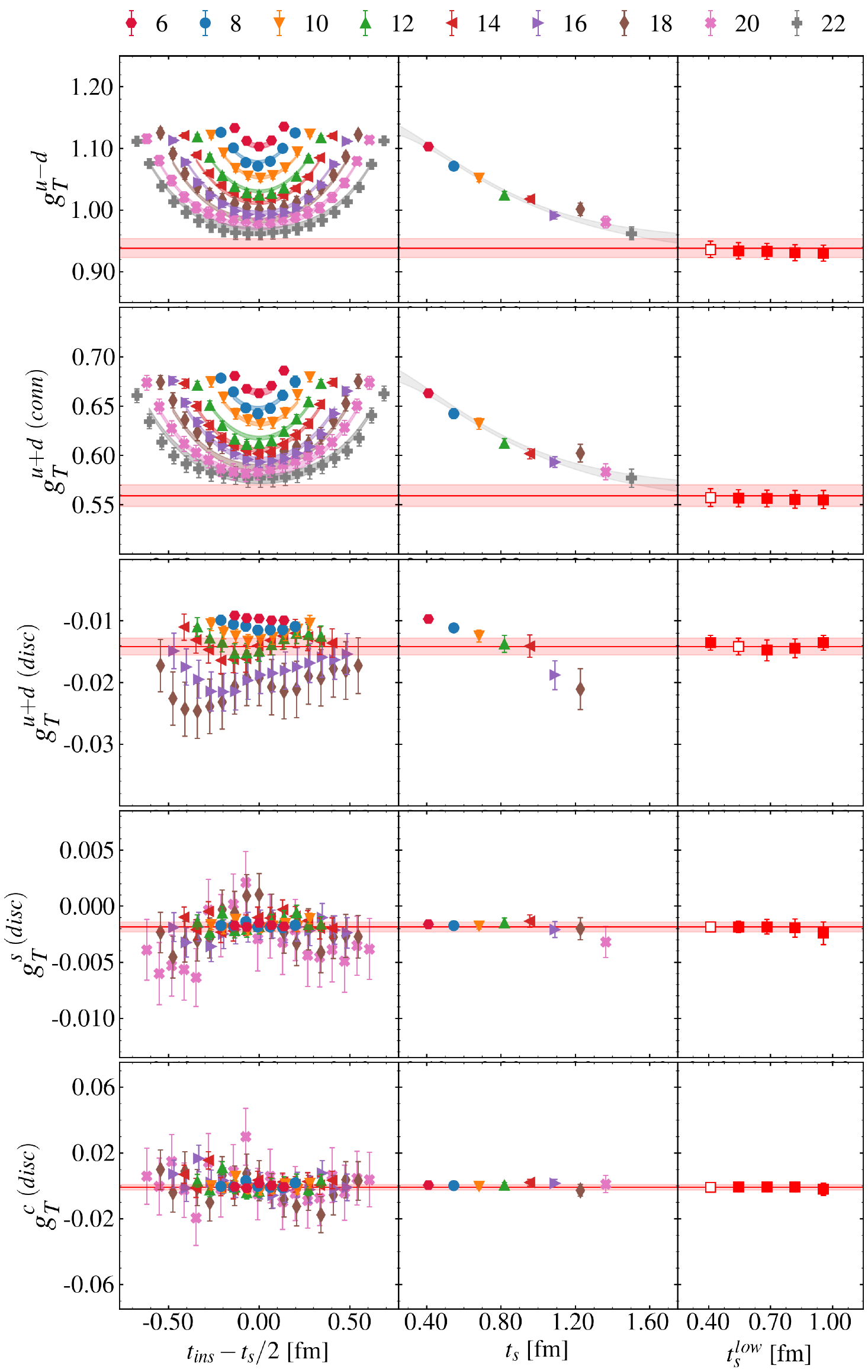}\\
	\vspace*{-0.3cm}
    \caption{The ratio and fit results for the C80 ensemble for tensor
      charges. The notation is the same as in Fig. \ref{fig:gA_B64}.}
	\vspace*{-0.4cm}
    \label{fig:gT_C80}
\end{figure}

\begin{figure}[h!]
	\includegraphics[width=\linewidth]{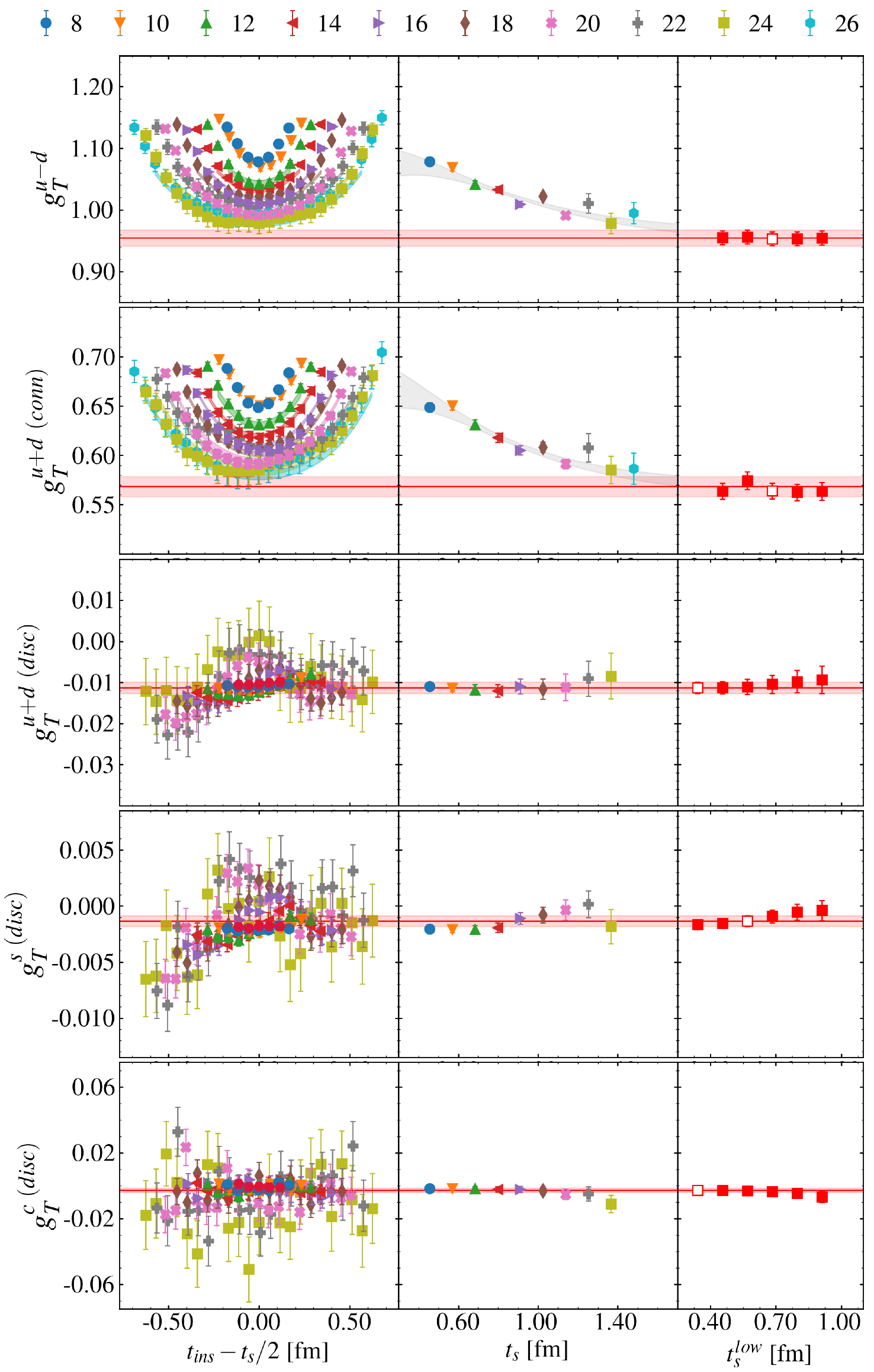}\\
	\vspace*{-0.3cm}
    \caption{The the ratio and fit results for the D96 ensemble for
      tensor charges. The notation is the same as in
      Fig. \ref{fig:gA_B64}.}
	\vspace*{-0.4cm}
    \label{fig:gT_D96}
\end{figure}

\begin{figure}[h!]
	\includegraphics[width=\linewidth]{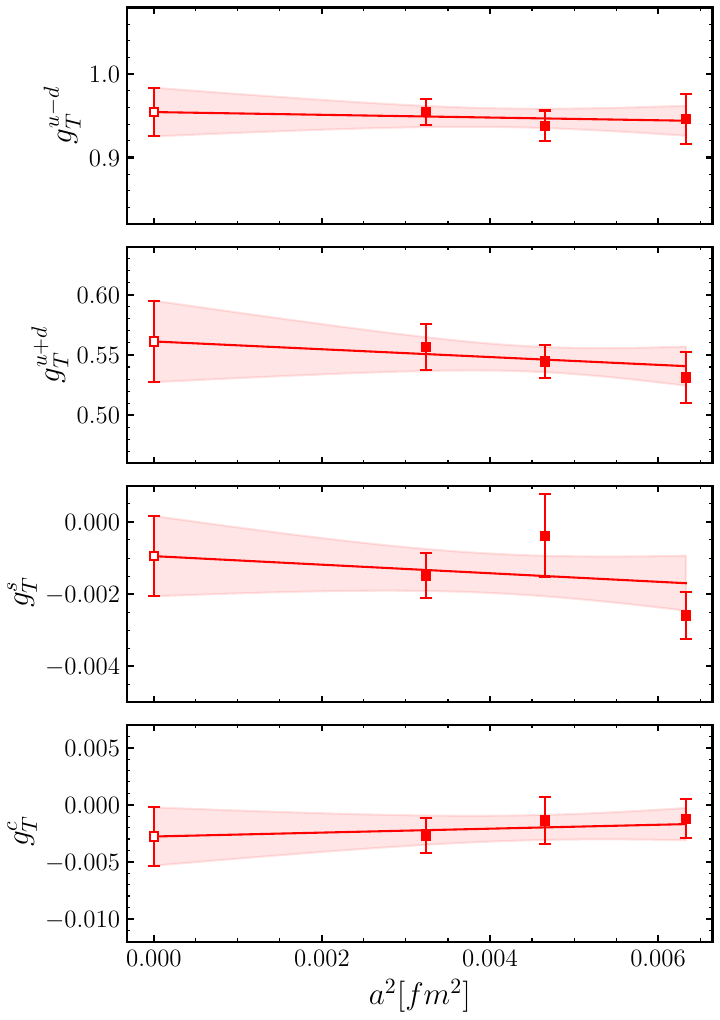}\\
	\vspace*{-0.3cm}
	\caption{Continuum limit of the nucleon tensor charges
          following the notation and procedure described in
          Fig.~\ref{fig:gA_extrap} for the axial case.}
	\vspace*{-0.4cm}
 \label{fig:gT_extrap}
\end{figure}

The same analysis is carried out for the tensor charges, shown in
Figs.~\ref{fig:gT_B64}, \ref{fig:gT_C80} and \ref{fig:gT_D96}, where,
similarly to the axial case, we observe excited state for the
connected contributions but not for the disconnected. We thus perform
2-state fits for the connected and plateau fits for the
disconnected. The continuum extrapolation in Fig.~\ref{fig:gT_extrap}
follows the same procedure as for the axial and scalar cases presented
so far.  The values of the tensor charges are presented in
Table~\ref{tbl:gT}.

\begin{table}[ht!]
	\centering
	{\footnotesize
		\renewcommand{\arraystretch}{1.2}
		\renewcommand{\tabcolsep}{2.0pt} 
		\begin{tabular}{c|c|c|c|c|c}
        \hline \hline
        $g_T$  & $u-d$     & $u+d$      & $u+d-2s$    & $u+d+s-3c$   & $u+d+s+c$   \\ \hline
        B64    & 0.947(30) & 0.531(21)  & 0.537(21)   & 0.533(21)    & 0.528(21)   \\ 
        C80    & 0.938(18) & 0.545(14)  & 0.544(13)   & 0.547(14)    & 0.541(13)   \\ 
        D96    & 0.955(15) & 0.557(19)  & 0.560(19)   & 0.563(20)    & 0.553(20)   \\ \hline
        $a= 0$ & 0.955(29) & 0.561(34)  & 0.561(33)   & 0.569(37)    & 0.557(34)   \\ \hline \hline
		\end{tabular}
        \\[0.1cm]
		\begin{tabular}{c|c|c|c|c}
        \hline \hline
        $g_T$  & $u$       & $d$         & $s$          & $c$          \\ \hline
        B64    & 0.739(23) & -0.207(11)  & -0.00259(64) & -0.0012(17)   \\ 
        C80    & 0.741(14) & -0.1976(73) & -0.0004(12)  & -0.0014(21)  \\ 
        D96    & 0.756(16) & -0.1989(72) & -0.00149(62) & -0.0027(16)  \\ \hline
        $a= 0$ & 0.756(29) & -0.196(12)  & -0.0009(11)  & -0.0028(26)  \\ \hline \hline
		\end{tabular}
	}
	\caption{Values for the tensor 2-, 3-, and 4-flavor isovector
          and isoscalar combinations (top) and the extracted single
          flavor charges (bottom) for each ensemble and in the
          continuum limit.}
	\label{tbl:gT}
	\vspace*{-0.2cm}
\end{table}

\subsection{Nucleon $\sigma$-terms}

The nucleon $\sigma$-terms are defined as 
\begin{gather}
    \sigma^f= m_f \braket{N|\bar{\psi}_f  \psi_f|N}~,~~\sigma^{u+d}=m_{ud} \braket{N|\bar{u}u+\bar{d}d|N},
\end{gather}
for a given quark $\psi_f$ of flavor $f$, where $m_f$ is the quark
mass, or for the isoscalar combination with $m_{ud}$ the average light
quark mass and $\ket{N}$ the nucleon state. The value of
$\sigma^{u+d}$ or $\sigma^{\pi N}$ is also known from phenomenological
analyses using input from experiment. These quantities are fundamental
in QCD as they give the quark content of the nucleon and their values
are a measure of chiral symmetry breaking.

We extract the nucleon $\sigma$-terms, using the matrix elements of
the scalar operator directly, including the disconnected quark
loops. In the twisted mass formulation the renormalization is simpler
compared to standard Wilson, since there is no additive mass
renormalization for the quark masses and the multiplicative
renormalization of the scalar current and of the quark mass
cancel. Our results for $\sigma^{\pi N}$, $\sigma^s$ and $\sigma^c$
are given in Table~\ref{tbl:sigma_terms}.

\begin{table}[ht!]
	\centering
	{\normalsize
		\renewcommand{\arraystretch}{1.2}
		\renewcommand{\tabcolsep}{3.0pt}
		\begin{tabular}{c|c|c|c}
        \hline \hline
               & $\sigma^{\pi N}$ & $\sigma^{s}$ & $\sigma^{c}$ \\ \hline
        B64    & 50.1(7.2)      & 56(12)       & 126(43)      \\ 
        C80    & 47.4(4.8)      & 45.4(7.6)    & 43(41)       \\ 
        D96    & 43.7(3.6)      & 37.5(6.0)    & 80(110)      \\ \hline 
        $a= 0$ & 41.9(8.1)      & 30(17)       & 82(29)       \\ \hline \hline
        \end{tabular}
	}
	\caption{Results for the  nucleon $\sigma$-terms for each ensemble and in the continuum. For $\sigma_{\pi N}$ and  for $\sigma_s$ we follow the same extrapolation procedure as described in Fig.~\ref{fig:gA_extrap}, while for $\sigma_c$ we use a single constant extrapolation.}
	\label{tbl:sigma_terms}
	\vspace*{-0.2cm}
\end{table}

\subsection{Comparison with other results}

\begin{figure}[htbp]
    \centering
    \begin{minipage}{\columnwidth}
        \centering
        \includegraphics[width=0.8\linewidth]{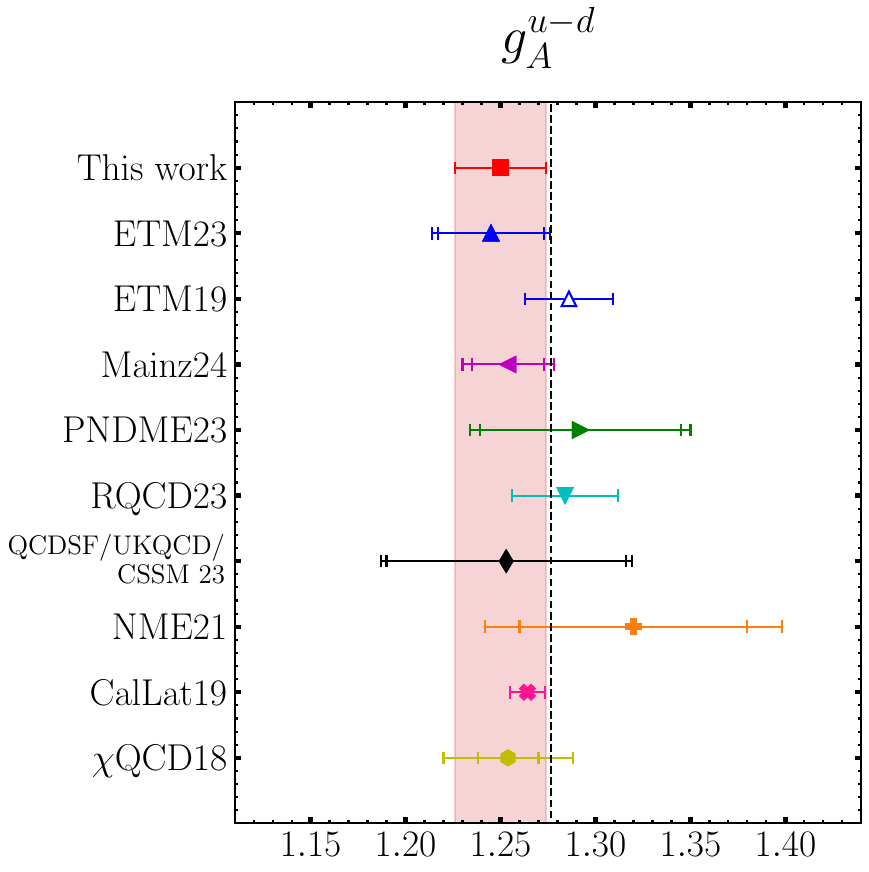}
        \vspace{0.5em}
        \includegraphics[width=0.8\linewidth]{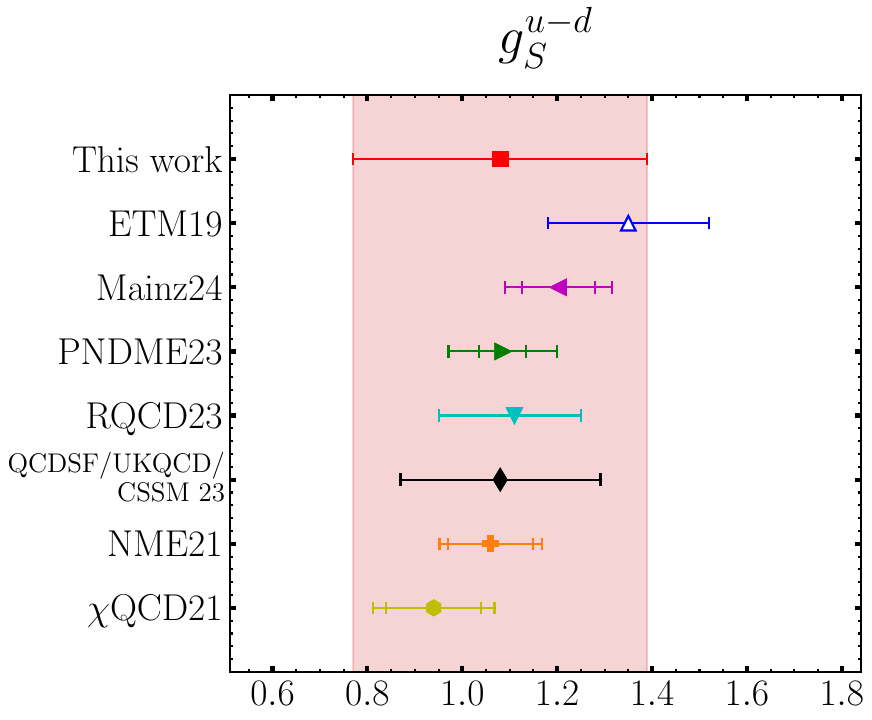}
        \vspace{0.5em}
        \includegraphics[width=0.8\linewidth]{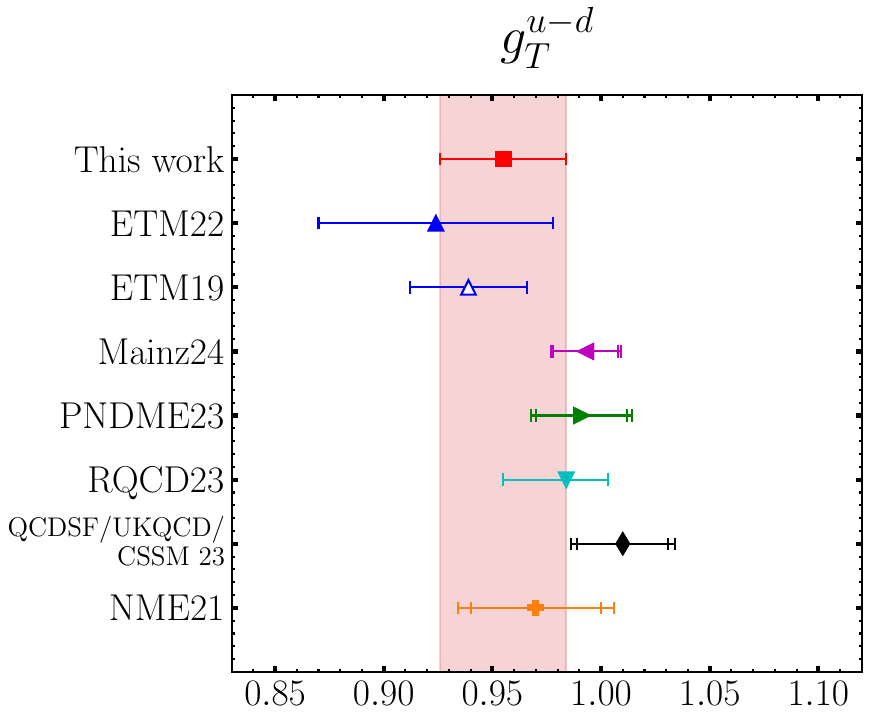}
        \caption{Comparison of the results of this work with other lattice QCD results, for the isovector axial (top), scalar (center), and tensor (bottom) charges. Our results are shown with the red square and red error band. The blue triangles show previous ETMC results \cite{Alexandrou:2023qbg,Alexandrou:2019brg,Alexandrou:2022dtc}. Open symbols represent results without a continuum limit extrapolation. The magenta triangles show recent results from the Mainz group~\cite{Djukanovic:2024krw}, the green triangles from PNDME~\cite{Jang:2023zts}, the cyan triangles from RQCD~\cite{Bali:2023sdi}, the black diamonds from the QCDSF/UKQCD/CSSM collaboration~\cite{QCDSFUKQCDCSSM:2023qlx}, the orange crosses from NME~\cite{Park:2021ypf}, the pink cross from CalLat~\cite{Walker-Loud:2019cif} and the yellow hexagons from $\chi$QCD~\cite{Liang:2018pis,Liu:2021irg}. For $\gaumd$, the dashed line represents the experimental value~\cite{Markisch:2018ndu}. }
        \label{fig:umd_comb}
    \end{minipage}
\end{figure}

\begin{figure}[htbp]
	\includegraphics[width=\linewidth]{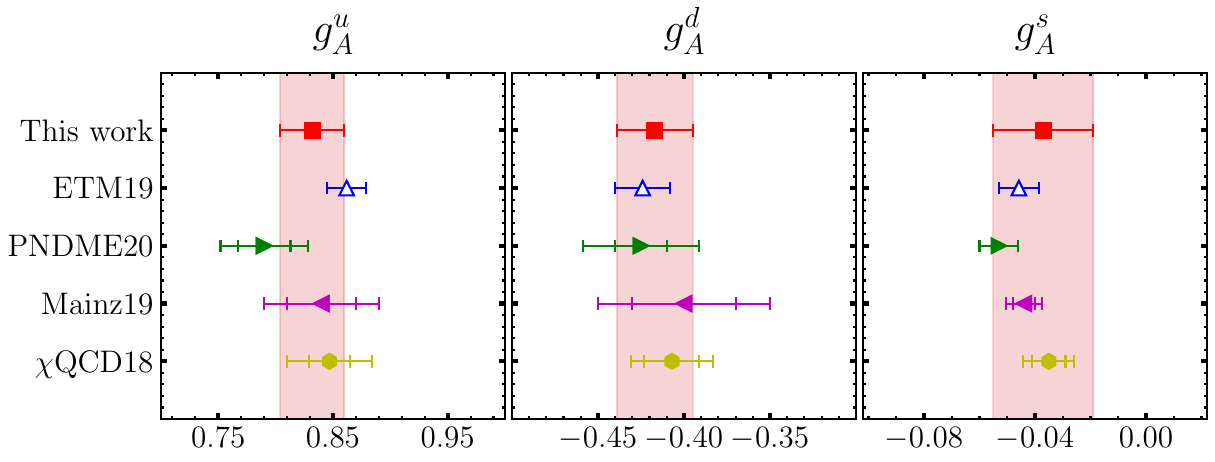}\\
	\vspace*{-0.3cm}
	\caption{Comparison of our results for the up- (left), down- (center), and strange-quark (right) contributions to the  axial charge with other lattice QCD  studies. Our results are shown with the red square and red error band. The blue triangles show previous ETMC results \cite{Alexandrou:2019brg}. The green triangles show results from PNDME~\cite{Park:2020axe}, the magenta triangles from the Mainz group~\cite{Djukanovic:2019gvi} and the yellow hexagons from $\chi$QCD~\cite{Liang:2018pis}.  }
	\vspace*{-0.4cm}
        \label{fig:gA_flavs}
\end{figure}

\begin{figure}[htbp]
	\includegraphics[width=\linewidth]{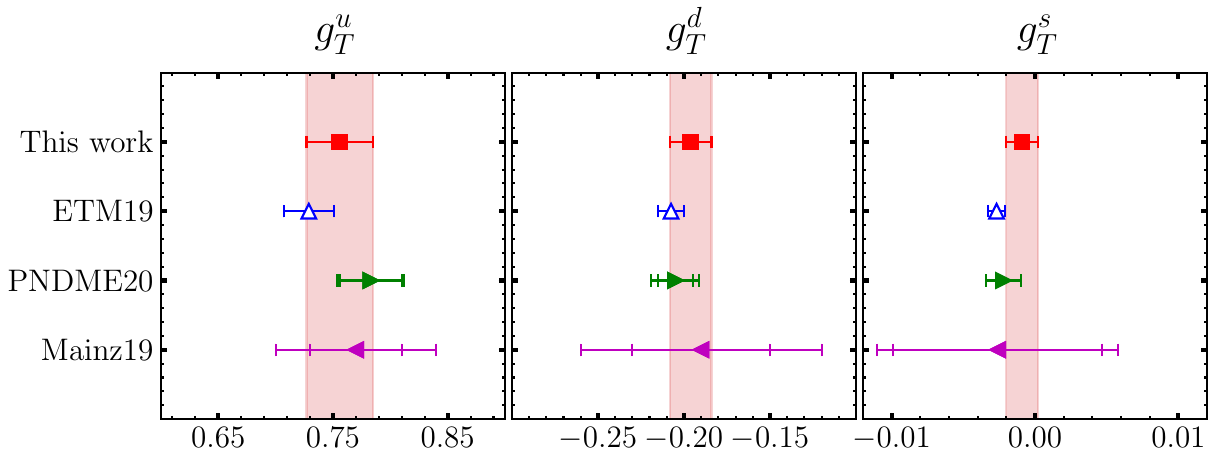}\\
	\vspace*{-0.3cm}
	\caption{Comparison of our results for the up- (left), down- (center), and strange-quark (right) contributions to the tensor charges with other lattice QCD analyses. Our results are shown with the red square and red error band. The blue triangles show previous ETMC results \cite{Alexandrou:2019brg}. The green triangles show results from PNDME~\cite{Park:2020axe} and the magenta triangles from the Mainz group~\cite{Djukanovic:2019gvi}.}
	\vspace*{-0.4cm}
        \label{fig:gT_flavs}
\end{figure}

\begin{figure}[htbp]
	\includegraphics[width=\linewidth]{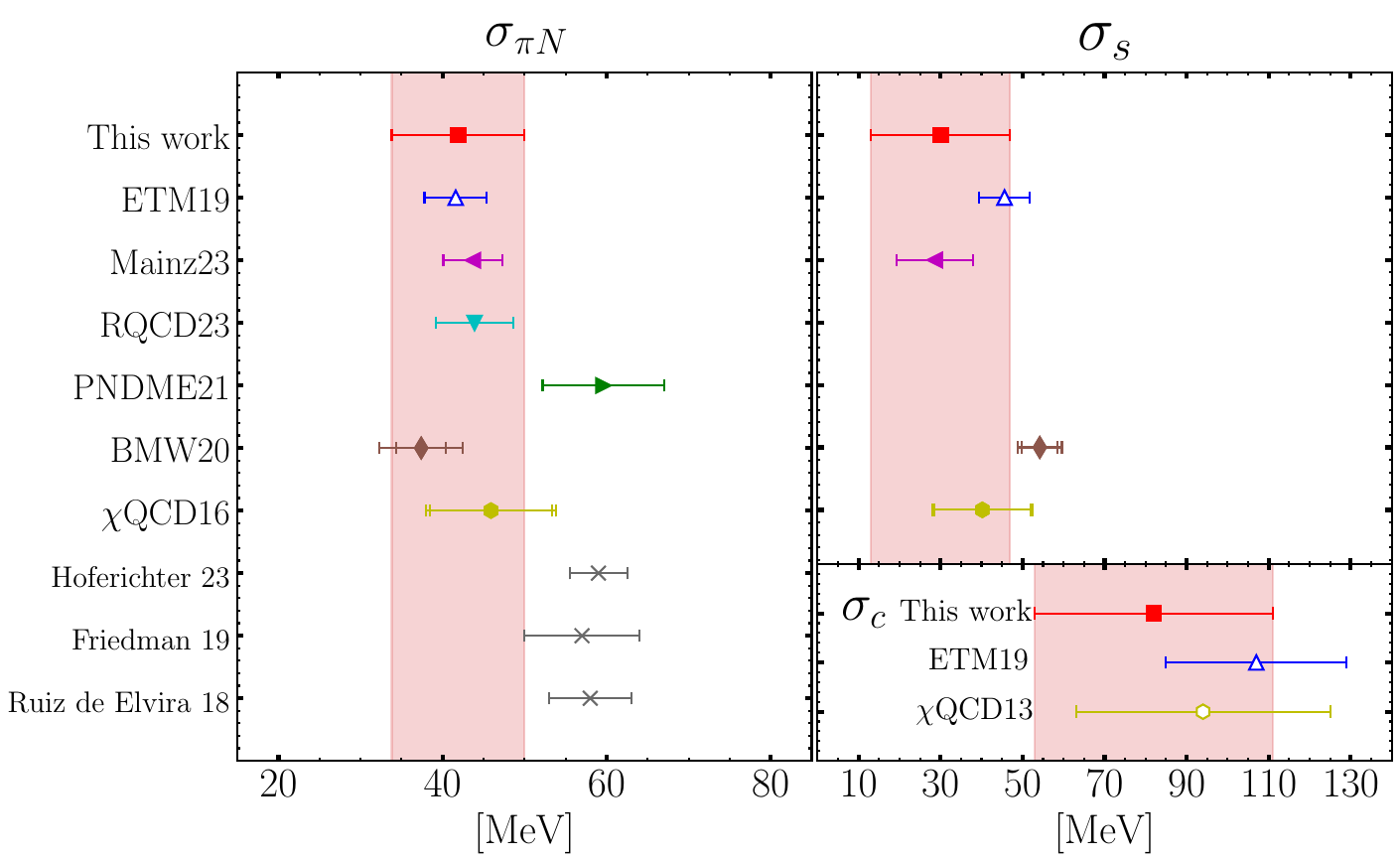}\\
	\vspace*{-0.3cm}
	\caption{Comparison of our results for the nucleon sigma terms with other lattice works and with results from phenomenology for $\sigma_{\pi N}$. Our results are shown with the red square and red error band. The blue triangles show previous ETMC results \cite{Alexandrou:2019brg}. The magenta triangles show results from the Mainz group~\cite{Agadjanov:2023efe}, the cyan triangles from RQCD~\cite{Bali:2023sdi}, the green triangles from PNDME~\cite{Gupta:2021ahb}, the brown diamonds from BMW~\cite{Borsanyi:2020bpd} and the yellow hexagons from $\chi$QCD~\cite{Yang:2015uis,XQCD:2013odc}. Open symbols represent results without a continuum limit extrapolation. Results from phenomenology as shown in grey crosses~\cite{Friedman:2019zhc,RuizdeElvira:2017stg,Hoferichter:2023ptl}.  }
	\vspace*{-0.4cm}
        \label{fig:sigma_comb}
\end{figure}

In this section we compare  our results with  recent results from other lattice QCD studies as well as from phenomenology. We first comment on previous analyses of these quantities  by ETMC:

\begin{itemize}
    \item 
    In Ref.~\cite{Alexandrou:2023qbg}, a comprehensive study of the
    nucleon isovector axial and pseudoscalar form factors is performed
    using the same three ensemble analyzed here, including a
    systematic error analysis. The final result for the isovector
    axial charge is obtained using only a single linear extrapolation
    compared to the model average strategy used in this work. We
    denote this result by ETM23 in Fig.~\ref{fig:umd_comb}.

    \item In Ref.~\cite{Alexandrou:2022dtc}, the isovector tensor
      charge was extracted using the same three ensembles analyzed
      here. In this work, we increase statistics for the B64 ensemble
      and for the two-point functions for the D96 ensemble. We also
      implement the AIC for extracting our final result.  We denote
      this result by ETM22 in Fig.~\ref{fig:umd_comb}.

    \item In Ref.~\cite{Alexandrou:2019brg}, an analysis of the same
      quantities presented here was carried out for the B64 ensemble,
      the only ensemble of the three available at the time. Thus,
      those results were obtained without a continuum limit
      extrapolation and in addition no model averaging was
      performed. We denote these results by ETM19 in
      Figs.~\ref{fig:umd_comb}-\ref{fig:sigma_comb}.
\end{itemize}

Comparing among ETMC results, we observe that for the isovector axial charge $\gaumd$ shown in Fig.~\ref{fig:umd_comb}, ETMC values are consistent and the errors are approximately the same after taking the continuum limit. For the isovector scalar charge $\gsumd$, we observe an increase in the error due to the continuum extrapolation. For the  isovector tensor charge $g_T^{u-d}$, there is only a slight increase in the error compared to the one obtained using only the B64 gauge ensemble. However, the current improved analysis yields a smaller error as compared to our  previous analysis using the same three ensembles and taking the continuum limit.

We also compare our results with those from other collaborations, noting that the results presented here are the only ones obtained with the continuum limit taken using ensembles simulated directly at the physical pion mass. In contrast, all collaborations to which we compare to here rely either exclusively on ensembles simulated with heavier-than-physical pion masses or combining one or two physical point ensembles with heavier-than-physical ensembles. Since results at the physical point carry larger statistical uncertainties in general, their extrapolations may be weighted more heavily by data from the heavier-than-physical ensembles rather than from their physical point simulations. Given that chiral extrapolations for nucleon quantities are less reliable compared to the meson sector~\cite{Shanahan:2016pla}, this approach may introduce unaccounted systematic errors, which our work avoids by not requiring such extrapolations.

In Figs. \ref{fig:umd_comb}, \ref{fig:gA_flavs} and \ref{fig:gT_flavs} we compare our results for the nucleon charges  with previous ETMC results and values from other collaborations and in Fig. \ref{fig:sigma_comb} we provide a similar comparison for  the $\pi N$, strange and charm $\sigma$-terms. 
Below we give some details on recent studies by other lattice QCD collaborations on the nucleon charges and $\sigma$-terms.

\begin{itemize}
    \item The PNDME collaboration  analyzed  thirteen $N_f = 2+1+1$ ensembles of the MILC collaboration, simulated using  highly improved staggered quarks (HISQ) in the sea, and using a mixed action approach whereby they use clover-improved valence quarks to compute the nucleon correlation functions. The gauge ensembles are  at four lattice spacings, 0.06~fm, 0.09~fm, 0.12~fm and 0.15 fm, three pion masses (135~MeV, 220~MeV and 310~MeV), and volumes with $3.7 \leq m_\pi L \leq 5.5$. Two of these ensembles are at the physical pion mass. A combined continuum limit  and chiral extrapolation was performed to extract the isovector nucleon charges. Their results are denoted as PNDME23~\cite{Jang:2023zts} in Fig.~\ref{fig:umd_comb}.
    
    The PNDME collaboration also extracted the up, down and strange  axial and tensor charges~\cite{Park:2020axe} using the same setup as for the isovector charges and nine $N_f = 2+1+1$ ensembles  with lattice spacings  from ~0.06 fm to ~0.15 fm, pion masses from 136 MeV to 320 MeV and volumes with $3.7 \leq m_\pi L \leq 4.79$. These results are denoted as PNDME20 in Figs.~\ref{fig:gA_flavs} and~\ref{fig:gT_flavs}.

   Furthermore, an evaluation of the $\sigma_{\pi N}$ was carried by PNDME~\cite{Gupta:2021ahb} using  six $N_f = 2+1+1$ ensembles at  three lattice spacings, 0.12,~0.09, and~0.06~fm, three pion mass values, with one ensemble at 138 MeV and the remaining at 230 MeV and 315 MeV, and volumes with $3.90 \leq m_\pi L \leq 4.79$. Their quoted result, denoted by PNDME21 in Fig.~\ref{fig:sigma_comb}, is obtained by explicitly treating $N\pi$ and $N\pi \pi$  excited states. This is motivated by chiral perturbation theory and brings their result in agreement with phenomenology. 

    \item The RQCD collaboration performed an extensive study of nucleon charges and $\sigma$-terms~\cite{Bali:2023sdi} using forty-seven $N_f = 2+1$ clover-improved fermion ensembles mostly produced  within the Coordinated Lattice Simulations (CLS) effort. These ensembles are simulated using pion masses spanning from 130 Mev to 430 MeV, with six lattice spacings between $a \approx 0.039 fm$ and $a \approx 0.098 fm$ and volumes with $3.0 \leq m_\pi L \leq 6.5$ with two of them having pion mass close to the physical value. They obtained their final values, denoted by RQCD23 in Figs.~\ref{fig:umd_comb} and~\ref{fig:sigma_comb}, after a simultaneous continuum and chiral extrapolation.

    \item The  Mainz collaboration for their isovector charges results in Ref.~\cite{Djukanovic:2024krw} used fifteen $N_f = 2+1$ CLS ensembles with pion masses in the range of 130 MeV $\leq m_\pi \leq$ 360 MeV, four lattice spacings and volumes with $3.05 \leq m_\pi L \leq 5.89$. They performed continuum and chiral extrapolations, with one ensemble having a pion mass close to its physical value and one about 30~MeV above. Their results on the isovector charges are denoted by Mainz24 in Fig.~\ref{fig:umd_comb}. In addition, they computed the up, down and strange  axial and tensor charges~\cite{Djukanovic:2019gvi}, using eight CLS $N_f = 2+1$ gauge ensembles, with four lattice spacings ranging from 0.050~fm to 0.086~fm and volumes with $3.78 \leq m_\pi L \leq 5.29$. In this analysis they have no physical point ensembles, using pion masses in the range 200~MeV-360 Mev. They carry out a combined fit to extrapolate to the continuum limit and to the physical pion mass obtaining the results denoted by Mainz19 in Figs.~\ref{fig:gA_flavs} and~\ref{fig:gT_flavs}.

    To extract the $\sigma_{\pi N}$ and $\sigma_s$,  sixteen CLS ensembles were analyzed with pion masses in the range 174~MeV $\leq m_\pi \leq$ 352 MeV, lattice spacings ranging from 0.050~fm to 0.086~fm and volumes with $3.00 \leq m_\pi L \leq 5.83$~\cite{Agadjanov:2023efe}. Continuum and chiral extrapolations are performed. The results are denoted by Mainz23 in Fig.~\ref{fig:sigma_comb}.

    \item The  QCDSF/UKQCD/CSSM collaboration~\cite{QCDSFUKQCDCSSM:2023qlx} used twenty-one  $N_f = 2+1$ gauge ensembles, a tree-level Symanzik improved gluon action and nonperturbatively $\ord(a)$ clover-improved Wilson fermions. Their pion masses are in the range 220 MeV $\leq m_\pi \leq$ 468 MeV, with five lattice spacings and three lattice volumes. They performed continuum and chiral extrapolations to extract the isovector charges shown in  Fig.~\ref{fig:umd_comb}.

    \item The NME collaboration, for their calculation of the isovector charges~\cite{Park:2021ypf}, used seven $N_f = 2+1$ clover-improved Wilson fermion ensembles. Their pion masses are in the range 166 MeV $\leq m_\pi \leq$ 285 MeV, with lattice spacings ranging from 0.071~fm to 0.127~fm and volumes with $3.75 \leq m_\pi L \leq 6.15$. Results are obtained after chiral and continuum limit extrapolations. These results are denoted as NME21 in Fig.~\ref{fig:umd_comb}.

    \item The CalLat collaboration computed  $\gaumd$ using an approach inspired by the  Feynman-Hellmann Theorem~\cite{Chang:2018uxx, Walker-Loud:2019cif}. They use  a hybrid setup of $N_f = 2+1+1$ HISQ staggered fermion in the sea generated by the MILC collaboration  and domain-wall fermions as valence quarks.   Sixteen gauge ensembles  were analyzed, spanning three lattice spacings ($a \approx 0.15,~0.12,~0.09$ fm) and pion mass values in the range 130-400 MeV. The final value, denoted as CalLat19 in Fig.~\ref{fig:umd_comb}, is obtained after a simultaneous chiral and continuum extrapolation.

    \item The $\chi$QCD collaboration computed the isovector and the up, down and strange axial charges~\cite{Liang:2018pis} using a hybrid setup of domain wall fermion sea and overlap valence quarks. They analyzed  three RBC/UKQCD $N_f = 2 + 1$ gauge ensembles with three lattice spacings, $a=0.08$ fm, $a=0.11$ fm and $a=0.14$ fm, two volumes and three pion masses 171~MeV, 302~MeV and 337 MeV. They performed continuum and chiral extrapolations and the results are denoted by $\chi$QCD18 in Figs.~\ref{fig:umd_comb} and \ref{fig:gA_flavs}.

    The $\chi$QCD collaboration also computed  $\gsumd$, denoted as $\chi$QCD21 in Fig.~\ref{fig:umd_comb}. They analyzed five RBC/UKQCD $Nf = 2 + 1$ domain-wall ensembles, with four lattice spacings, $a=0.06, 0.08, 0.11,0.14$ fm~\cite{Liu:2021irg}, one ensemble with physical pion mass and the rest with  pion masses in the range 171~MeV to 371~MeV at three different volumes. 
    
    For the calculation of the pion-nucleon and strange $\sigma$-terms~\cite{Yang:2015uis}, three RBC/UKQCD $Nf = 2 + 1$ domain-wall ensembles were used with lattice spacings, $a=0.08$ fm and $a=0.11$ fm, three pion masses, 139~MeV, 300~MeV and 330~Mev, and three different volumes. Their results on $\sigma_{\pi N}$ and $\sigma_s$ are obtained after chiral and continuum extrapolations and are denoted by $\chi$QCD16 in Fig.~\ref{fig:sigma_comb}.
    $\chi$QCD is the only collaboration, besides ETMC, that computed the charm $\sigma$-term by directly evaluating the charm quark loop. The calculation of $\sigma_c$~\cite{XQCD:2013odc} was done using only one ensemble with pion mass 331 MeV, volume $24^3 \times 64$ and $a=0.11$fm. For all other quantities we only compare to other collaborations when the results are  at the physical mass point and in the continuum limit. However, since this is the only other lattice QCD result on $\sigma_c$ we include it in Fig.~\ref{fig:sigma_comb} denoted as $\chi$QCD13.

    \item The  BMW collaboration used  the Feynman-Hellmann relation based on the derivative of the nucleon mass  with respect to the corresponding quark mass  to determine $\sigma_{\pi N}$, $\sigma_s$ and $\sigma_c$~\cite{Borsanyi:2020bpd}.  For that purpose, they used 33 $N_f = 1 + 1 + 1 + 1$ 3HEX-smeared, clover-improved Wilson ensembles with pion masses in the range of 195~MeV to 420 MeV, the strange and charm quark masses spanning their physical values, four lattice spacings in the range $0.06-0.10$~fm and spatial volumes $3.7 \leq m_\pi L \leq 12.2$. Their values for $\sigma_{\pi N}$ and $\sigma_s$, denoted as BMW20, are shown in Fig.~\ref{fig:sigma_comb}. Their error on $\sigma_c$ is very large and thus not included in the comparison.
    
\end{itemize}

In Fig. \ref{fig:umd_comb}, where we show results for the isovector charges, we observe very good agreement among lattice QCD results by different collaborations for all isovector charges. Moreover, our value for the isovector axial charge $\gaumd$, is compatible with experimental measurements~\cite{Markisch:2018ndu}.
As can be seen in Fig. \ref{fig:gA_flavs} and \ref{fig:gT_flavs}, there is a good agreement among lattice QCD results for the up, down, strange and charm  axial and tensor charges.

In Fig. \ref{fig:sigma_comb}, where we collected  results for the $\pi N$, strange and charm $\sigma$-terms we observe that for the isovector scalar charge, there is a significant error increase when taking the continuum limit as compared to our previous results using only the B64 gauge ensemble~\cite{Alexandrou:2019brg}.  While there is an   overall good agreement among lattice results, a tension of about two standard deviations is observed with the latest value from  the PNDME collaboration~\cite{Gupta:2021ahb}, where the pion-nucleon contribution was explicitly included using chiral perturbation theory, yielding a value that is closer to phenomenological results. The BMW collaboration used a different method to extract the $\sigma$-terms computing the dependence of the nucleon mass on the quark masses. Their values agree with the ones extracted from the the computation of the nucleon three-point function where excited states may contribute differently, giving confidence on the proper extraction of the nucleon matrix element. Furthermore, in our recent study~\cite{Alexandrou:2024tin,Alexandrou:2023ajp}, where we included the pion-nucleon interpolating fields in a generalized eigenvalue problem analysis, we found no detectable improvement. This indicates  that the tension of lattice QCD results on $\sigma_{N\pi}$ with phenomenology has a different origin.

\section{Conclusions}
\label{sec:conclusions}

We present results on the nucleon axial, scalar and tensor charges, as well as the nucleon $\sigma$-terms, using three $N_f=2+1+1$ ensembles with twisted mass clover-improved fermions, with quark masses tuned to reproduce their physical values. This enables us, for the first time, to determine these charges and $\sigma$-terms  at the continuum limit  using only physical point ensembles, avoiding any chiral extrapolations.

We find for the isovector charges:
\begin{gather}
    g_A^{u-d}=1.250(24),~g_S^{u-d}=1.08(31),~g_T^{u-d}=0.955(29), \notag
\end{gather}
that are in agreement with results by other collaborations. The value we get for the isovector axial charge also agrees with the experimental value. The scalar and tensor isovector charges determined here can provide  input for experimental efforts on  scalar and tensor interactions for BSM physics searches. 

We also use the scalar matrix element to extract the values of the $\sigma$-terms, that are of significant importance in the direct detection of dark matter. Our value $\sigma_{\pi N}=41.9(8.1)$, is in agreement with most lattice results, but within approximately one standard deviation  with phenomenological results that tend to give larger values.

The observed trend of increasing errors in several quantities, when taking the continuum limit, emphasizes the significance of utilizing more physical point ensembles  preferably with smaller lattice spacings in taking the continuum limit. Our goal in the  future is the inclusion of such an additional  ensemble with a$\sim 0.05$ for taking the continuum limit, increasing the accuracy in the final values.

\section*{ACKNOWLEDGMENTS}
We thank all members of the ETM collaboration for a most conducive
cooperation.  C.A. and G. K. acknowledge partial support from the
European Joint Doctorate AQTIVATE that received funding from the
European Union's research and innovation programme under the Marie
Sklodowska-Curie Doctoral Networks action under the Grant Agreement No
101072344. Y.L. is supported by the Excellence Hub project "Unraveling
the 3D parton structure of the nucleon with lattice QCD (3D-nucleon)"
id EXCELLENCE/0421/0043 co-financed by the European Regional
Development Fund and the Republic of Cyprus through the Research and
Innovation Foundation and by funding provided by the University of
Cyprus through the project "Nucleon generalized parton distributions
within lattice QCD (Nucleon-GPDs)". Ch. I. is supported by the project
Nucleon-GPDs. S.~B. and G.~K. acknowledge support by the Excellence
Hub project "NiceQuarks" id EXCELLENCE/0421/0195 co-financed by the
European Regional Development Fund S.B.~and J.F.~acknowledge financial
support from the Inno4scale project, which received funding from the
European High-Performance Computing Joint Undertaking (JU) under Grant
Agreement No.~101118139. The JU receives support from the European
Union's Horizon Europe Programme.  J.F.~received support by the DFG
research unit FOR5269 ``Future methods for studying confined gluons in
QCD'' and acknowledges financial support by the Next Generation
Triggers project (https://nextgentriggers.web.cern.ch).

The open-source packages tmLQCD~\cite{Jansen:2009xp,Abdel-Rehim:2013wba,Deuzeman:2013xaa,Kostrzewa:2022hsv}, LEMON~\cite{Deuzeman:2011wz}, DD-$\alpha$AMG~\cite{Frommer:2013fsa,Alexandrou:2016izb,Bacchio:2017pcp,Alexandrou:2018wiv}, QPhiX~\cite{joo2016optimizing,Schrock:2015gik} and QUDA~\cite{Clark:2009wm,Babich:2011np,Clark:2016rdz} have been used in the ensemble generation. The open-source software PLEGMA has been used for the analyais.

The authors gratefully acknowledge the Gauss Centre for Supercomputing e.V.~(www.gauss-centre.eu) for providing computing time on the GCS Supercomputers SuperMUC-NG at Leibniz Supercomputing Centre, JUWELS~\cite{JUWELS} and JUWELS Booster~\cite{JUWELS-BOOSTER} at Juelich Supercomputing Centre (JSC).  Part of the results were created within the EA program of JUWELS Booster also with the help of the JUWELS Booster Project Team (JSC, Atos, ParTec, NVIDIA). 
We acknowledge the Swiss National Supercomputing Centre (CSCS) and the EuroHPC Joint Undertaking for awarding this project access to the LUMI supercomputer, owned by the EuroHPC Joint Undertaking, hosted by CSC (Finland) and the LUMI consortium through the Chronos programme under project IDs CH16-CYP.
We are grateful to CINECA and the EuroHPC JU
for awarding  access to  
supercomputing facilities hosted at CINECA and  
to Leonardo-Booster, provided to us through the 
Extreme Scale Access Call grant EHPC-EXT-2024E01-027.
The authors also acknowledge computing time granted on Cyclone at the Cyprus institute (CyI) via the project with ids P061, P146 and pro22a10951.

\bibliography{refs}

\end{document}